\documentclass[journal ]{new-aiaa}
\usepackage[utf8]{inputenc}
\usepackage{textcomp}

\usepackage{graphicx}
\usepackage{amsmath}
\usepackage[version=4]{mhchem}
\usepackage{siunitx}
\usepackage{subfig}
\usepackage{tikz}
\usepackage{listings}
\usepackage{longtable,tabularx}
\usepackage{color}
\usepackage{colortbl}
\usepackage{algorithm}
\usepackage{algpseudocode,setspace}
\usepackage{xspace}
\usepackage{hyperref}
\hypersetup{colorlinks=true,linkcolor=blue,filecolor=magenta,urlcolor=cyan}

\newcommand{\mc}[1]{\multicolumn{1}{c}{#1}}

\title{A numerical investigation of sweep effects on turbulent flow over iced wings}

\author{Ziyu Zhou\footnote{Master Student, School of Aerospace Engineering, Tsinghua University. Email: zhouziyu23@mails.tsinghua.edu.cn}}
\affil{School of Aerospace Engineering, Tsinghua University, MasterBeijing 100084, China}
\author{Maochao Xiao\footnote{Ph.D. Student, Department of Mechanical and Aerospace Engineering, Sapienza University of Rome. Email: maochao.xiao@uniroma1.it}}
\affil{Department of Mechanical and Aerospace Engineering, Sapienza University of Rome, Rome 00184, Italy}
\author{Jiawei Chen\footnote{Postdoctoral Researcher, Department of Mathematics, University of Texas at Arlington. Email: chenjiaweidr@163.com}}
\affil{Department of Mathematics, University of Texas at Arlington, Arlington 76019, USA}
\author{Li Li\footnote{Corresponding author. Professor, Department of Computer Science, Northwestern Polytechnical University; Researcher, AVIC Xi’an Aeronautics Computing Technique Research Institute. Email: westlili@163.com}}
\affil{Department of Computer Science, Northwestern Polytechnical University, Xi’an 710065, China\\
AVIC Xi´an Aeronautics Computing Technique Research Institute, Shaanxi 710000, China}
\author{Yufei Zhang\footnote{Corresponding author. Associate Professor, School of Aerospace Engineering, and State Key Laboratory of Advanced Space Propulsion, Tsinghua University. Email: zhangyufei@tsinghua.edu.cn}}
\affil{School of Aerospace Engineering, Tsinghua University, Beijing 100084, China\\
State Key Laboratory of Advanced Space Propulsion, Tsinghua University, 100084, Beijing}

\begin{document}

\maketitle

\begin{abstract}
This study employs an improved delayed detached eddy simulation (AMD-IDDES) method to investigate the flow characteristics of an iced swept wing. The AMD-IDDES approach incorporates an anisotropic minimum-dissipation (AMD) model into the traditional IDDES framework to better account for local flow anisotropy and energy dissipation characteristics. The method is validated against the traditional IDDES approach, and the results show that AMD-IDDES more accurately captures the evolution of vortices and separated shear layers. Compared with an iced straight wing, a swept wing exhibits distinct aerodynamic behavior driven mainly by the sweep angle. The sweep angle induces strong spanwise flow, which reshapes the separation region and transforms the flow from two-dimensional to three-dimensional. This spanwise motion significantly alters vortex development and enhances the complexity of the unsteady flow. The shear layer separation, initiated by a Kelvin–Helmholtz instability, dominates the unsteady aerodynamic response. Although wingtip effects remain secondary, their interaction with the leading-edge vortex still contributes to lowering the dominant frequency. Overall, this study highlights that the aerodynamic forces of an iced swept wing are primarily governed by sweep-induced spanwise flow and associated shear layer dynamics, with minimal influence from the root and tip regions.
\end{abstract}

\section{Introduction}\label{sec:intro}
When an aircraft traverses clouds containing supercooled water droplets, it experiences icing as those droplets freeze upon contact. Icing disrupts the aircraft's aerodynamic shape, leading to premature boundary layer transition and early flow separation \cite{liNumericalSimulationIced2021}. Ice accretions on aircraft surfaces can severely degrade aerodynamics, causing decreases in lift, increases in drag, and the deterioration of force moments \cite{braggIceairfoilAerodynamics2005}. Such ice accretions are categorized based on their geometry, with horn ice on an airfoil leading edge being one of the most dangerous \cite{stebbinsReviewComputationalMethods2019a}. This type of ice induces a separated shear layer due to the large concave curvature behind the horn, triggering a Kelvin--Helmholtz (K--H) instability and secondary instabilities in the shear layer and promoting the transition to fully developed turbulence.\par
Before 2000, Reynolds-averaged Navier–Stokes (RANS) simulations were used exclusively to model these phenomena. Such simulations proved inadequate for accurately predicting mean aerodynamic loads or reproducing flow unsteadiness. RANS methods typically underestimate lift coefficients by generating elongated ice-induced separation bubbles, largely due to their inability to model the nonequilibrium turbulence in the ice-induced separated shear layer \cite{xiaoEnhancedPredictionThreedimensional2022}. This limitation becomes increasingly significant on approaching stall conditions \cite{liNumericalSimulationIced2021}. Consequently, attention later shifted toward hybrid RANS/large eddy simulation (RANS/LES) methods, and detached eddy simulation (DES) 
and its variants gained prominence \cite{braggAirfoilIceAccretionAerodynamic2007}. The subsequent development of hybrid methods and the continuous increase in computational resources mean that the potential for hybrid methods in the aerospace industry is growing \cite{stebbinsReviewComputationalMethods2019a}, particularly for more complex geometries.\par
Most simulations for iced airfoils/wings employ hybrid RANS/LES methods due to their favorable balance between accuracy and computational cost, particularly in high-Reynolds-number scenarios where LES can be prohibitively expensive. Chen et al. \cite{chenNumericalStudyEffects2024} modified the $\gamma\text{-}Re_{\theta}$ transition model and enhanced its predictive accuracy for complex flows over discontinuous three-dimensional ice shapes. RANS-based approaches may struggle to capture the unsteady nature of separated flows \cite{ansellMeasurementUnsteadyFlow2014, ozcerNumericalStudyIced2023a}, especially under large-scale flow separations induced by horn-shaped ice formations. Thus, further research is needed to enhance the accuracy of these methods and deepen our understanding of complex flow behaviors over 3D iced wings, especially for high angles of attack.\par
There has been considerable previous research on the flow fields of iced airfoils, including studies on the GLC305 \cite{zhangZonalDetachedEddySimulation2015,molinaApplicationDDESIced2020a,xiaoImprovedPredictionFlow2021}, NACA 0012 \cite{butlerImprovedDelayedDetachedEddy2016}, NLF 0414 \cite{xiaoNumericalStudyIced2019}, and 30P30N \cite{xiaoNumericalInvestigationUnsteady2020,leeLargeeddySimulationsComplex2020} airfoils. However, research on wings (which have a limited span compared to two-dimensional airfoils) is still relatively scarce. Xiao et al. \cite{xiaoEnhancedPredictionThreedimensional2022} used shear-layer adaptive improved delayed detached-eddy simulation (SLA-IDDES) to simulate the flow over a NACA 0012 iced wing with a finite span.
Broeren et al. \cite{broerenLowReynoldsNumberAerodynamics2017a} conducted simulations on the $65\%$ scale Common Research Model (CRM65) wing, incorporating leading-edge ice shapes derived from NASA experiments. Bornhoft et al. \cite{bornhoftLargeeddySimulationsCRM652024a,bornhoftUseArtificialIce2025} and Stebbins et al. \cite{stebbinsNumericalSimulationIced2024} further investigated related flow phenomena. Additionally, recent WMLES studies by Craig Penner et al. \cite{craigpennerWallModeledLargeEddySimulations2024b,craigpennerWallModeledSweptWing2025b} on the CRM65 wing with leading-edge ice are also recognized. Therefore, this study aims to simulate more complex ice configurations and further investigate the impact of ice accretion on flow fields. As a second goal, this study also seeks to develop and employ more accurate and robust computational methods.\par

Zhou et al. \cite{zhouEnhancedDelayedDetachededdy2025} developed a turbulence model based on the anisotropic minimum-dissipation (AMD) subgrid length scale model \cite{rozemaMinimumdissipationModelsLargeeddy2015}, leading to the creation of the AMD-IDDES method. It has been demonstrated that this model effectively addresses anisotropic grids and significantly alleviates the “gray area” issue, which refers to the delayed formation of Kelvin–Helmholtz instabilities in separated shear layers due to inadequate resolution and excessive modeled dissipation near the RANS–LES interface. Consequently, this study aims to simulate ice accretion on a finite aspect ratio swept wing using both the IDDES and the improved AMD-IDDES methods. In \S\ref{sec:numerics}, the AMD-IDDES method is introduced and validated using decaying isotropic turbulence scenarios to ensure its reliability in turbulence modeling. \S\ref{sec:Computional setup} details the computational setup. The aerodynamic results are analyzed in \S\ref{sec:Force and Pressure Coefficients}, \S\ref{sec:Vortex Structures and Flow Characteristics}, and \S\ref{sec:Pressure Fluctuation PSD Analysis}. The force and pressure coefficient distributions, vortex structures, and flow characteristics of the iced swept wing are presented. The pressure fluctuation power spectral density (PSD) and proper orthogonal decomposition (POD) are used to identify characteristic frequencies. Finally, \S\ref{sec:conclusions} summarizes the key findings and discusses potential directions for future research.

\section{Computational methodology}\label{sec:numerics}

\subsection{Improved delayed detached-eddy simulation with anisotropic minimum-dissipation subgrid length scale}\label{sec:AMD-IDDES}
The AMD-IDDES method is an improved version of the IDDES approach designed to address the “gray area” issue inherent in the standard IDDES model, thereby enhancing the accuracy of simulations for complex flow scenarios, especially separated and transitional flows. This method combines the advantages of the standard IDDES framework with an AMD model, which adjusts the subgrid length scale based on local flow characteristics. The AMD subgrid length scale was proposed by Rozema et al. \cite{rozemaMinimumdissipationModelsLargeeddy2015} to provide the minimum eddy dissipation required to dissipate the energy of subfilter scales. The model locally approximates the exact dissipation and is consistent with the nonlinear gradient model. By incorporating these improvements into the DES framework, the AMD-IDDES method enhances the prediction of key flow features such as separation bubbles, vortex shedding, and reattachment zones, providing more accurate results for flows with strong anisotropy. \par
The SST $k-\omega$ shear-stress transport model, which serves as the baseline RANS formulation for AMD-IDDES, is described in \cite{gritskevichDevelopmentDDESIDDES2012}. The method consists of replacing the RANS length scale in the destruction term of the turbulent kinetic energy (TKE) equation with an IDDES hybrid length scale such that the model behaves as both RANS and LES in one simulation. The modified TKE equation is
\begin{equation}
  \frac{\partial(\rho k)}{\partial t}+\frac{\partial(\rho u_jk)}{\partial x_j}=P_k-\frac{\rho k^{3/2}}{l_{IDDES}}+\frac{\partial}{\partial x_j}\Big[(\mu+\sigma_k\mu_t)\frac{\partial k}{\partial x_j}\Big], \label{eqn:TKE}
\end{equation}
where $l_{IDDES}$ is the IDDES hybrid length scale defined as
\begin{equation}
    l_{IDDES}=f_{d}(1+f_{e})l_{RANS}+(1-f_{d})l_{LES},
\end{equation}
and $l_{_{RANS}}=\sqrt{k} / (C_{_\mu}\omega$) and $l_{LES}$ are the RANS turbulence length scale and the LES subgrid length scale, respectively. The term $f_d$ is a blending function ranging between 0.0 (LES mode) and 1.0 (RANS mode). The LES subgrid scale $l_{LES}$ is given by
\begin{equation}
    l_{LES}=\min\{l_{wall},l_{free}\}
\end{equation}
where $l_{wall} = C_w \max [d_w,\Delta_{\max} ]$ and $l_{free}$ are the LES subgrid length scales in the near vicinity of the walls and the regions free from wall effects, respectively. $C_w$ is a constant, $d_w$ is the distance to the closest wall, and $\Delta_{\max}$ is the maximum edge length of a cell.\par
In AMD-IDDES, $l_{free}$ is taken as the AMD subgrid length scale. The eddy viscosity of the AMD model is expressed as \cite{zhouEnhancedDelayedDetachededdy2025} \par
\begin{equation}
    {\Delta _{AMD}} = {\left( {\frac{{ - {\Delta _k}{g_{ik}}{\Delta _k}{g_{jk}}{S_{ij}}}}{{{g_{ml}}{g_{ml}}}}} \right)^{\frac{1}{2}}},
\end{equation}
\begin{equation}
    C_{DES,AMD} = C_A \left( \frac{\gamma}{\beta} \right)^{\frac{3}{4}},
\end{equation}
\begin{equation}
    l_{free} = C_{DES,AMD} \Delta_{AMD},
\end{equation}
where $c_A$ is a constant, $g_{ij}=\partial u_i / \partial x_j$, and $S_{ij}=(g_{ij}+g_{ji})/2$. Mathematically, $\Delta _kg_{ik}$ is the scaled velocity gradient. Rozema suggested that $C_A^2=0.30$ is a good choice for a second-order central difference scheme \cite{rozemaMinimumdissipationModelsLargeeddy2015}. This value yields $C_{DES,AMD}=1.92$ with $\gamma=0.44$ and $\beta=0.0828$. In our simulations, the parameter is calibrated as $C_{DES,AMD}=2.40$ using decaying isotropic turbulence following Zhou et al. \cite{zhouEnhancedDelayedDetachededdy2025}. \par
Furthermore, a lower limit is set for the length scale to restrict and ensure smaller LES length scales when the “gray area" issue arises. In the RANS-modeled attached boundary layers, the LES length scale could be so small that the hybrid length scale becomes less than the RANS length scale. Consequently, the RANS functionality could be contaminated and yield lower wall friction values. To circumvent these two possible defects, the LES length scale is restricted as \par
\begin{equation}
    \begin{array}{l}
    {l_{free} = \left( {{C_{DES,AMD}}{\Delta _{AMD}}} \right)_{\lim }}\\
     = \left\{ \begin{array}{l}
    \max \left( {{C_{DES,AMD}}{\Delta _{AMD}},{C_{\lim }}{V^{1/3}}} \right), \quad {f_d} \leq \varepsilon \\
    {C_{DES}}{\Delta _{\max }}, \quad {f_d} \ge \varepsilon 
    \end{array} \right.
    \end{array}
\end{equation}
where $C_{DES}$ is the parameter in the standard IDDES method, V is the cell volume, and the threshold value $\epsilon$ is set as 0.01. The ultimate LES length scale is
\begin{equation}
    l_{LES}={\min} \{ C_w \max [d_w,\Delta_{\max} ],(C_{DES,AMD} \Delta_{AMD})_{\lim} \}.
\end{equation}

Unlike the AMD-IDDES method, the standard IDDES method uses the maximum cell spacing scale for the subgrid scale, where $l_{free}=C_{DES} \Delta_{max} = C_{DES} \max(\Delta_{x},\Delta_{y},\Delta_{z})$ and $\Delta_x$, $\Delta_y$, and $\Delta_z$ are the grid scales along the three directions. \par

The Navier-Stokes equations are solved using CFL3D \cite{CFL3DVersion}, into which the AMD-IDDES model was implemented by the authors. This is a structured solver that uses a cell-centered finite volume method. Time integration is performed using a dual-time-stepping approximate factorization scheme \cite{klokerFullyImplicitScheme1986}, where subiterations and a multigrid approach are used for second-order accuracy and faster pseudo-time convergence. The viscous flux is approximated using a second-order central difference scheme. The inviscid flux is computed via a hybrid central/upwind scheme,
\begin{equation}
    F_{inviscid}=(1-\sigma)F_{central}+\sigma F_{upwind},
\end{equation}
where the central flux is approximated by the fourth-order central difference scheme, and the upwind flux is computed using the Roe scheme coupled with the third-order monotonic upstream-centered scheme for conservation laws (MUSCL) scheme \cite{xiaoEnhancedPredictionThreedimensional2022}. The blending function reads as 
\begin{equation}
    \sigma=\max\Bigg(\tanh\Bigg(\frac{C_3}{1-C_4}\max\Bigg(\frac{l_{IDDES}}{l_{RANS}}-C_4,0\Bigg)\Bigg),\sigma_{\min}\Bigg),
\end{equation}
where $C_3=4.0$ and $C_4=6.0$ \cite{liuDDESAdaptiveCoefficient2018}. In the separated regions, the ratio $l_{IDDES}/l_{RANS}$ is typically less than 0.6 \cite{xiaoImprovedPredictionFlow2021}, which leads to $\sigma=\sigma_{\min}$. $\sigma_{\min}=0.1$ in the present study.

\subsection{Calibration}\label{sec:Calibration and Test}
Decaying isotropic turbulence (DIT) is employed to calibrate the model coefficients and test the performance of IDDES and AMD-IDDES. The DIT is simulated in a cubic domain with periodic boundary conditions, corresponding to the experimental setup of Comte-Bellot and Corrsin (CBC) \cite{comte-bellotSimpleEulerianTime1971}. The experimental Reynolds number is $Re_M=U_0M/\nu=34000$, where $Re_M$ uses the grid width $M$ as the characteristic length and the uniform flow velocity $U_0$ as the characteristic velocity. Turbulent kinetic energy spectra are measured throughout the decay process. Our simulation starts from $U_0t/M=42$ and ends at $U_0t/M=98$. The computational model follows Zhou et al. \cite{zhouEnhancedDelayedDetachededdy2025}.\par
\begin{figure}[hbt!]
  \centering
  \includegraphics[width=0.55\columnwidth]{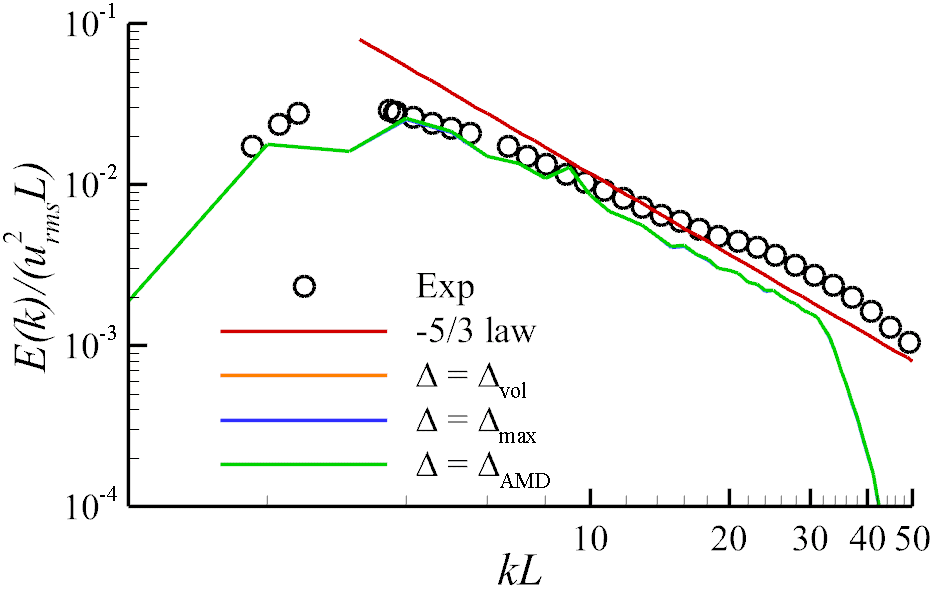}\hfill
  \caption{The energy spectrum on a $64\times64\times64$ grid.} 
  \label{fig:spectrum}
\end{figure}

Figure \ref{fig:spectrum} illustrates the computed spectra from various turbulence models on a $64\times64\times64$ grid, employing three definitions of the subgrid length scale: $\Delta = \Delta_{vol}$, $\Delta = \Delta_{max}$, and $\Delta = \Delta_{AMD}$. The results are obtained using $C_{DES}=0.61$ for IDDES and SLA-IDDES and $C_{DES,AMD}=2.40$ for AMD-IDDES. These values can yield results that fit both the experimental data and the theoretical $-5/3$ power law well. No pileup or excessive dissipation occurs in high-wavenumber ranges. On isotropic grids, all three subgrid length scales correspond to the cell size, which aligns perfectly with the overlap of the three energy spectra, indicating that the current parameter calibration is successful.\par
\begin{figure}[!t]
  \centering
  \subfloat[$32\times128\times32$]{\includegraphics[width=0.49\columnwidth]{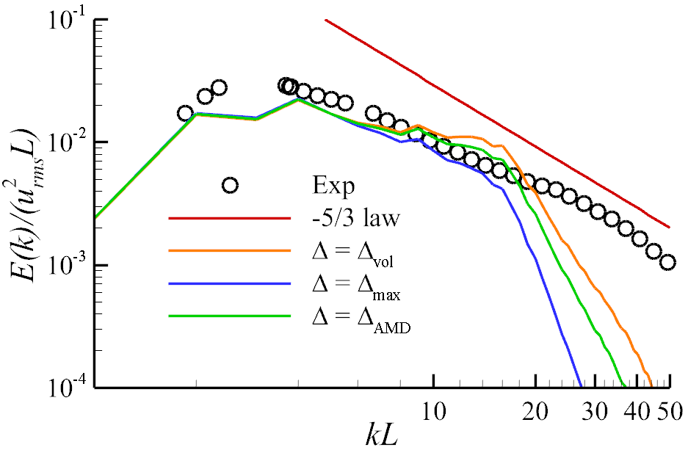}}%
  \label{DIT32.128.32}
  \hfil
  \subfloat[$128\times128\times32$]{\includegraphics[width=0.49\columnwidth]{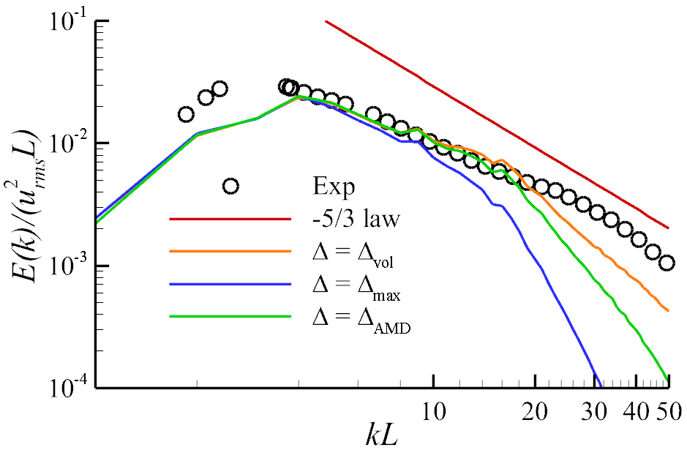}}%
  \label{DIT128.128.32}
  \caption{Energy spectra on different grids.}
  \label{fig:spectra2}
\end{figure}

Figure \ref{fig:spectra2} compares the results obtained using different definitions of the subgrid length scale on two types of anisotropic grids: the book cell ($\Delta y \ll \Delta x \sim \Delta z$) and the pencil cell ($\Delta y \sim \Delta x \ll \Delta z$). The book cell is shown on a $32\times128\times32$ grid, and the pencil cell is shown on a $128\times128\times32$ grid. A significant discrepancy is observed for the pencil-type grid when using $\Delta = \Delta_{max}$, which results in excessive dissipation in the high wavenumber range. This aligns with the established understanding that $\Delta = \Delta_{max}$ is a conservative approximation of the subgrid length scale. It is unreasonable for the $128\times128\times32$ grid to utilize the same subgrid length scale as a $32\times32\times32$ grid. Conversely, $\Delta = \Delta_{vol}$ causes a pileup in the high wavenumber range on the book-type grid due to being overly sensitive to the minimum cell edge length, representing an aggressive approximation of the subgrid length scale. Consequently, neither $\Delta = \Delta_{max}$ nor $\Delta = \Delta_{vol}$ can reproduce the experimental results. In contrast, the result of $\Delta = \Delta_{AMD}$ lies between the results obtained using $\Delta = \Delta_{max}$ and $\Delta = \Delta_{vol}$. \par
Therefore, the subgrid model associated with the IDDES method leads to excessive dissipation, which is the fundamental cause of the “gray area" issue in IDDES. The length scale obtained from $\Delta_{max}$ in the boundary layer is larger, resulting in a higher model eddy viscosity, inhibiting flow development. In contrast, the AMD-IDDES method improves the subgrid model, ensuring that dissipation remains at the correct level even on grids with significant anisotropy, effectively mitigating the “gray area" issue.\par

\section{Computational setup}\label{sec:Computional setup}
\begin{figure}[!t]
\centering
\subfloat[Straight wing model.]{\includegraphics[width=0.49\columnwidth]{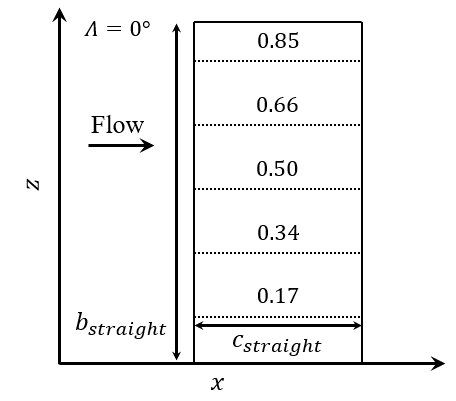}}
\label{model1}
\hfil
\subfloat[Swept wing model.]{\includegraphics[width=0.48\columnwidth]{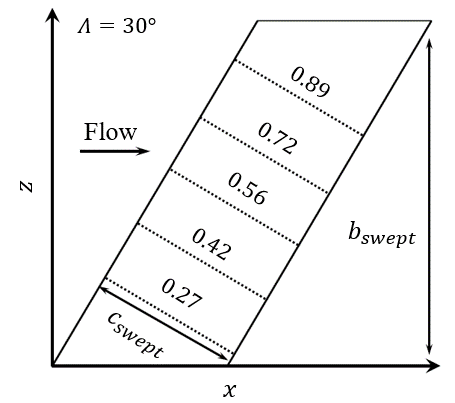}}
\label{model2}
\hfil \\
\subfloat[Ice model (adapted from Bragg et al. \cite{braggAerodynamicMeasurementsFinite1991}).]{\includegraphics[width=0.6\columnwidth]{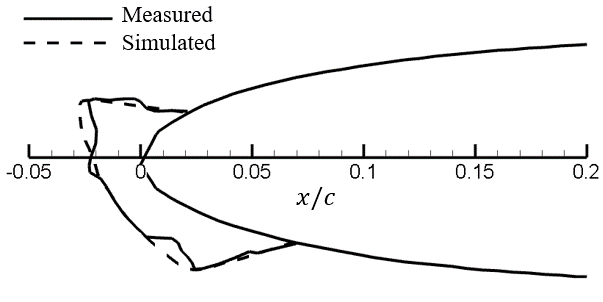}
\label{model3}}
\caption{Computational models for the iced straight and swept wings.}
\label{fig:model}
\end{figure}

The model used for the iced swept wing is a semi-span wing with a chord of 0.44 m and a span of 0.894 m. A NACA 0012 airfoil section (in the plane perpendicular to the leading edge) is used on this 30-degree swept wing. The icing conditions are a free-stream velocity of 58.12 m/s, an angle of attack of 4 degrees, an icing time of 5 minutes, volume median diameter droplet of 20 microns, $LWC=2.1$ g/m$^3$, and a temperature of $18~^\circ$F \cite{khodadoustAerodynamicsFiniteWing1995,braggEffectSimulatedIce1991}. The ice model distribution in $x$ corresponds to the chordwise direction rather than the flow direction. The model focuses on the data from five cross-sections, as shown in Figure \ref{fig:model}(b). These sections are perpendicular to the wing leading edge at $z/b = 0.27$, 0.42, 0.56, 0.72, and 0.89, respectively. \par
To analyze the flow characteristics of the iced swept wing in more detail, an iced straight wing is also simulated to highlight the unique features of the iced swept wing. The wing model is shown in Figure \ref{fig:model}(a), and the ice model distribution in $x$ corresponds to the flow direction. The computational model differs from the swept wing only in terms of the sweep angle. For details of the model, refer to Refs. \cite{khodadoustAerodynamicsFiniteWing1995,braggIceairfoilAerodynamics2005,xiaoEnhancedPredictionThreedimensional2022}. The sections are perpendicular to the wing leading edge and located at $z/b = 0.17$, 0.34, 0.50, 0.66, and 0.85.\par

\begin{figure}[!t]
  \centering
  \includegraphics[width=0.80\columnwidth]{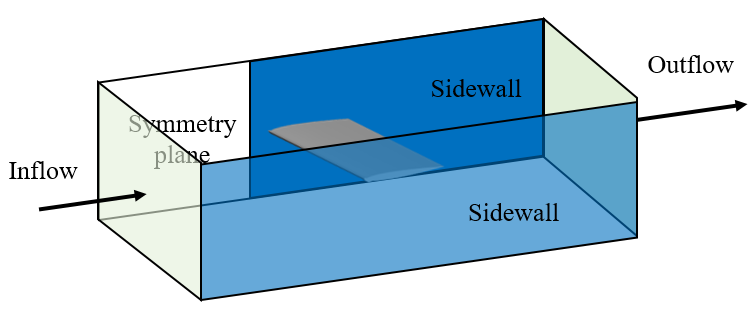}\hfill
  \caption{Computational domain for the iced swept wing (adapted from Li et al. \cite{liNumericalSimulationIced2021}).} 
  \label{fig:domian}
\end{figure}
\begin{figure}[!t]
\centering
\subfloat[The $x\text{-}y$ plane around the ice accretion.]{\includegraphics[width=0.5\columnwidth]{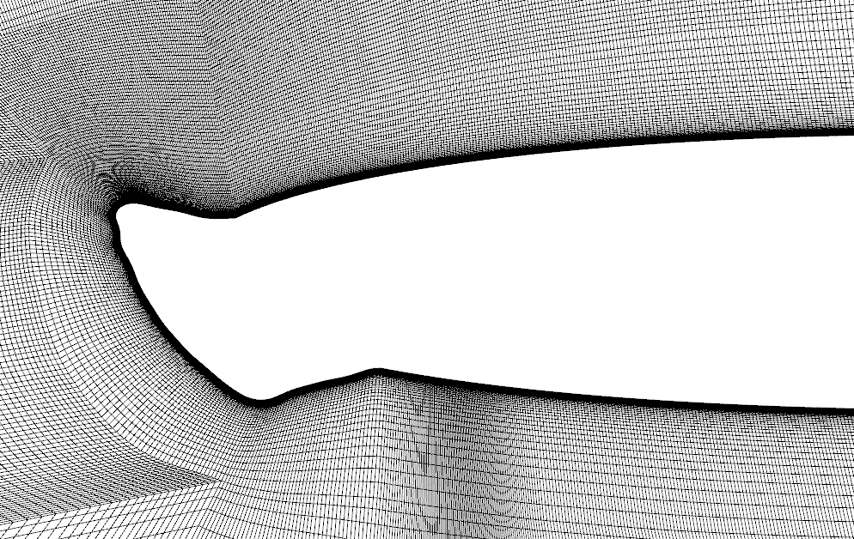}%
\label{grid1}}
\hfil
\subfloat[The $x\text{-}z$ plane.]{\includegraphics[width=0.45\columnwidth]{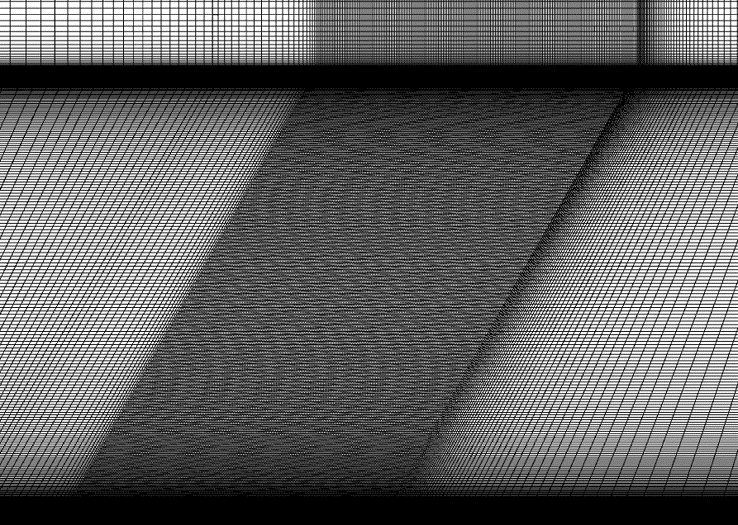}%
\label{grid2}}
\caption{The computational grid for the iced swept wing.}
\label{fig:grid}
\end{figure}

For the iced swept wing, the computational domain is the same as that in Li et al. \cite{liNumericalSimulationIced2021}, with a length of $3.3$ m along the flow direction, a width of $1.22$ m along the spanwise direction, and a height of $0.92$ m, as shown in Figure \ref{fig:domian}. The grid is shown in Figure \ref{fig:grid}. The mesh distribution for the iced straight wing is similar to that of the swept wing and thus is not presented separately. Table \ref{tbl:grid_d} describes the two sets of grids used in the study. $N_x$, $N_y$, and $N_z$ denote the cell numbers in the streamwise, normal, and spanwise directions, respectively. $\Delta x_{ice}/c$ is the streamwise grid spacing in the ice accretion region. The Reynolds number based on the chord length and inflow velocity is $Re=1.5 \times 10^6$, the Mach number is $Ma=0.2$, and the angle of attack is $AOA=8^\circ$. The inflow and outflow conditions are set as the inflow/outflow boundary conditions. The sidewalls are set with adiabatic nonslip conditions, but the wall where the model is installed is set as a slip wall in the inlet upstream of the model leading edge. The nondimensional physical time step is $\Delta tU_{\infty}/c=0.001$. The simulation was run for a total of 30 convective time units (CTUs), and the last 25 CTUs were used for statistical analysis after the flow had reached a fully developed state. For the iced straight wing, the flow and boundary conditions are exactly the same as for the iced swept wing. \par

\begin{table}[hbt!]
    \centering
    \caption{Grid details.}
    \label{tbl:grid_d}
    \begin{tabular}{c c c c c c c c}
      \hline
      \hline
      \mc{Grid}  & \mc{$N_x$} & \mc{$N_y$} & \mc{$N_z$} & \mc{$\Delta x_{ice}/c$} & \mc{$\Delta y_w /c$} & \mc{$\Delta z_w/c$} & \mc{$N_{total}$} \\
      \hline
      \small Straight, coarse  & $472$ & $176$ & $488$ & $0.0020$ & $5 \times 10^{-5}$ & $5 \times 10^{-5}$ & $56 \times 10^6$ \\
      \small Straight, fine    & $688$ & $252$ & $556$ & $0.0013$ & $5 \times 10^{-5}$ & $5 \times 10^{-5}$ & $119 \times 10^6$ \\
      \small Swept, coarse     & $532$ & $172$ & $476$ & $0.0020$ & $5 \times 10^{-5}$ & $5 \times 10^{-5}$ & $59 \times 10^6$ \\
      \small Swept, fine       & $756$ & $224$ & $552$ & $0.0013$ & $5 \times 10^{-5}$ & $5 \times 10^{-5}$ & $120 \times 10^6$ \\
      \hline
      \hline
    \end{tabular}
\end{table}
\section{Aerodynamic load characteristics}\label{sec:Force and Pressure Coefficients}
\begin{figure}[!t]
\centering
\subfloat[Lift.]{\includegraphics[width=0.32\columnwidth]{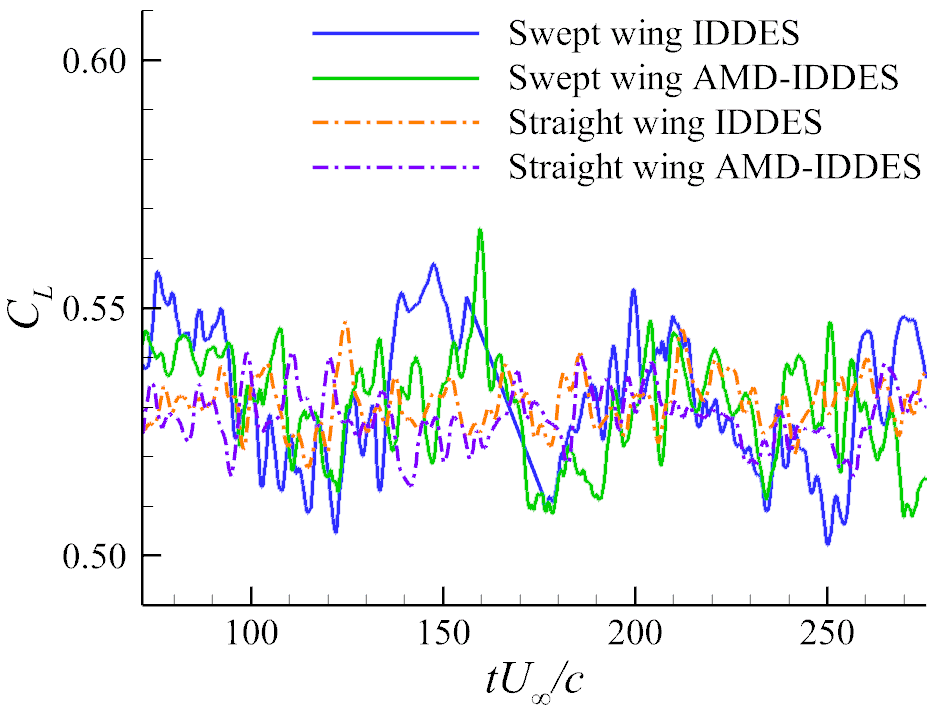}}%
\label{C-t1}
\hfil
\subfloat[Drag.]{\includegraphics[width=0.32\columnwidth]{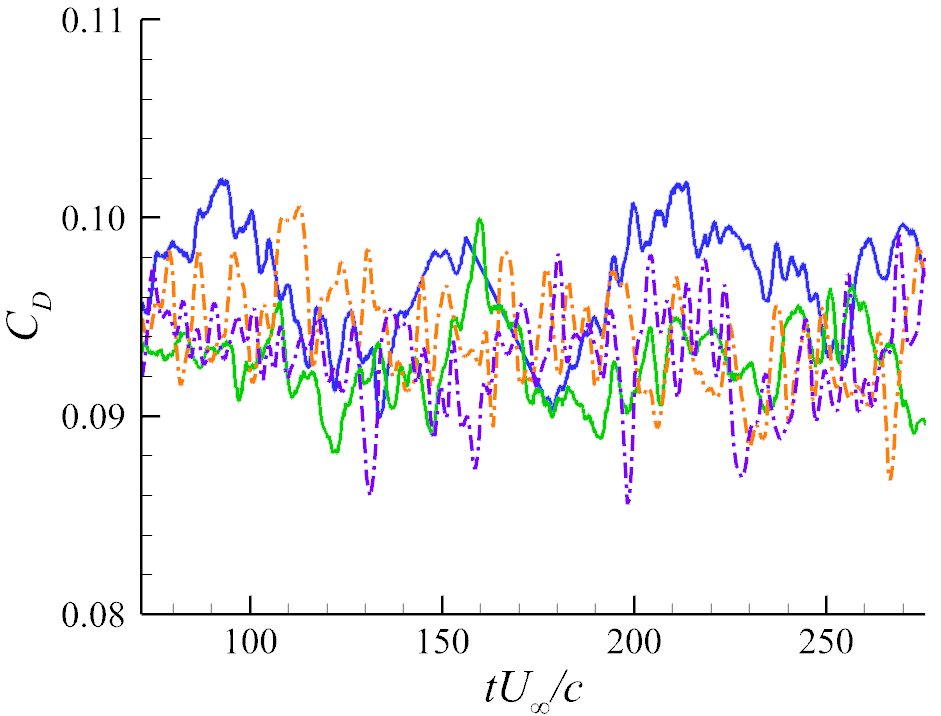}}%
\label{C-t2}
\subfloat[Pitching moment.]{\includegraphics[width=0.32\columnwidth]{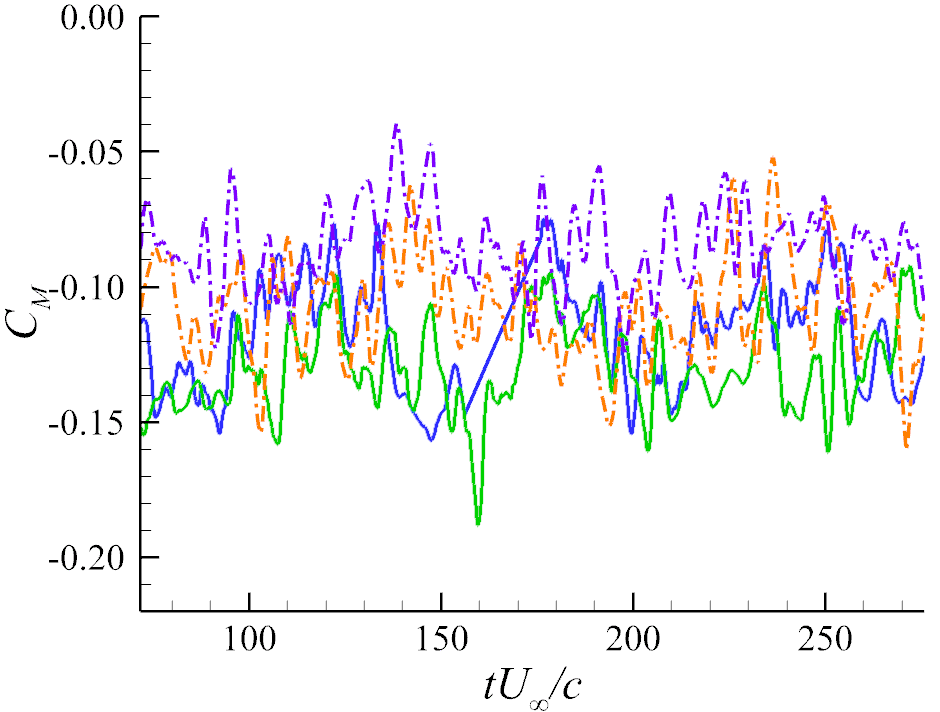}}%
\label{C-t3}
\caption{Integrated aerodynamic force (moment) coefficients.}
\label{fig:C-t}
\end{figure}

The temporal variations in the integrated forces and pitching moment on the fine grid are shown in Figure \ref{fig:C-t}, where the horizontal axis represents the nondimensional time used in CFL3D, defined as $\Delta t U_{\infty} /c$, with $U_{\infty}$ the freestream velocity, and $c$ the speed of sound. The observed trend and periodic distribution of the force coefficients indicate that the simulation has converged. Table \ref{tbl:C} lists the statistical results from IDDES and AMD-IDDES for the two wings. For both types of wings, the trends shown by the AMD-IDDES method are quite similar to those of the IDDES method. Both methods predict similar lift coefficients, but the average and root mean square values of all aerodynamic coefficients predicted by IDDES are slightly higher than those predicted by AMD-IDDES, especially for the drag coefficient. According to the drag polar plotted by Khodadoust et al. \cite{khodadoustAerodynamicsFiniteWing1995} for the midspan of the straight wing with and without simulated ice, the drag coefficient $C_D$ is approximately 0.097 at a lift coefficient $C_L$ of 0.526 for the iced wing. This value is in good agreement with the result predicted by the AMD-IDDES method in the present study, whereas the IDDES method overpredicts the drag coefficient significantly.\par
\begin{table}[hbt!]
  \centering
  \caption{Integrated lift, drag, and pitching moment.}
  \label{tbl:C}
  \begin{tabular}{ c c c c c c c }
    \hline
    \hline
    \mc{Method}    & \mc{$C_L$} & \mc{$C_{L,rms}$} & \mc{$C_D$} & \mc{$C_{D,rms}$} & \mc{$C_{M}$} & \mc{$C_{M,rms}$} \\
    \hline
    \small Straight, IDDES & $0.528$  & $0.0156$  & $0.094$  & $0.0069$ & $-0.088$  & $0.0067$  \\
    \small Straight, AMD-IDDES & $0.526$  & $0.0152$ & $0.093$  & $0.0064$  & $-0.079$  & $0.0050$ \\
    \small Swept, IDDES  & $0.530$  & $0.0130$  & $0.096$  & $0.0026$ & $-0.142$  & $0.0094$  \\
    \small Swept, AMD-IDDES & $0.528$  & $0.0106$ & $0.093$  & $0.0023$  & $-0.128$  & $0.0067$   \\
    \hline
    \hline
  \end{tabular}
\end{table}\par
\begin{figure}[!t]
\centering
\subfloat[Lift.]{\includegraphics[width=0.32\columnwidth]{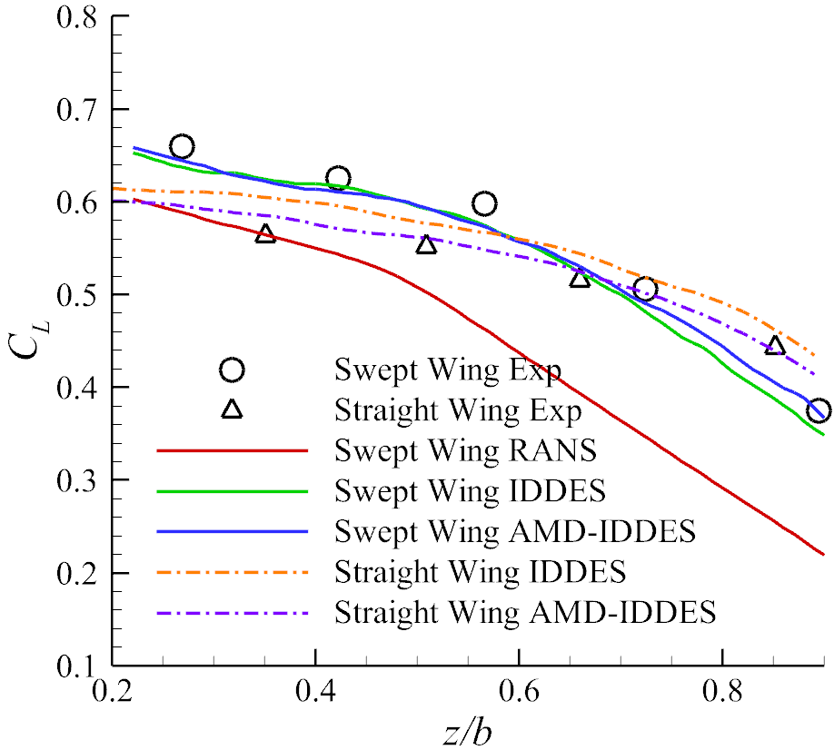}}%
\label{C-z1}
\hfil
\subfloat[Drag.]{\includegraphics[width=0.32\columnwidth]{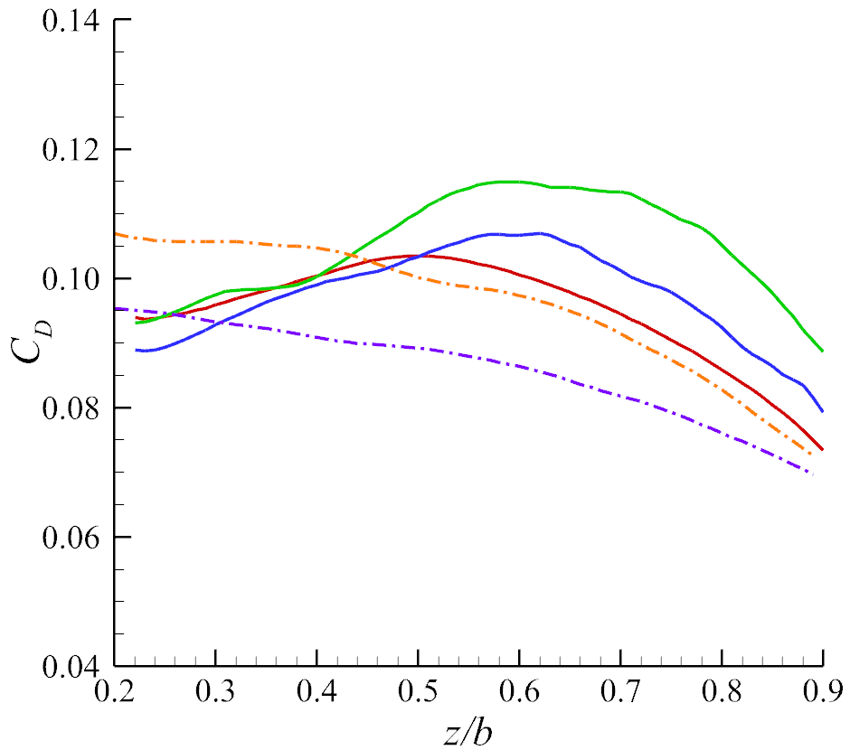}}%
\label{C-z2}
\subfloat[Pitching moment.]{\includegraphics[width=0.32\columnwidth]{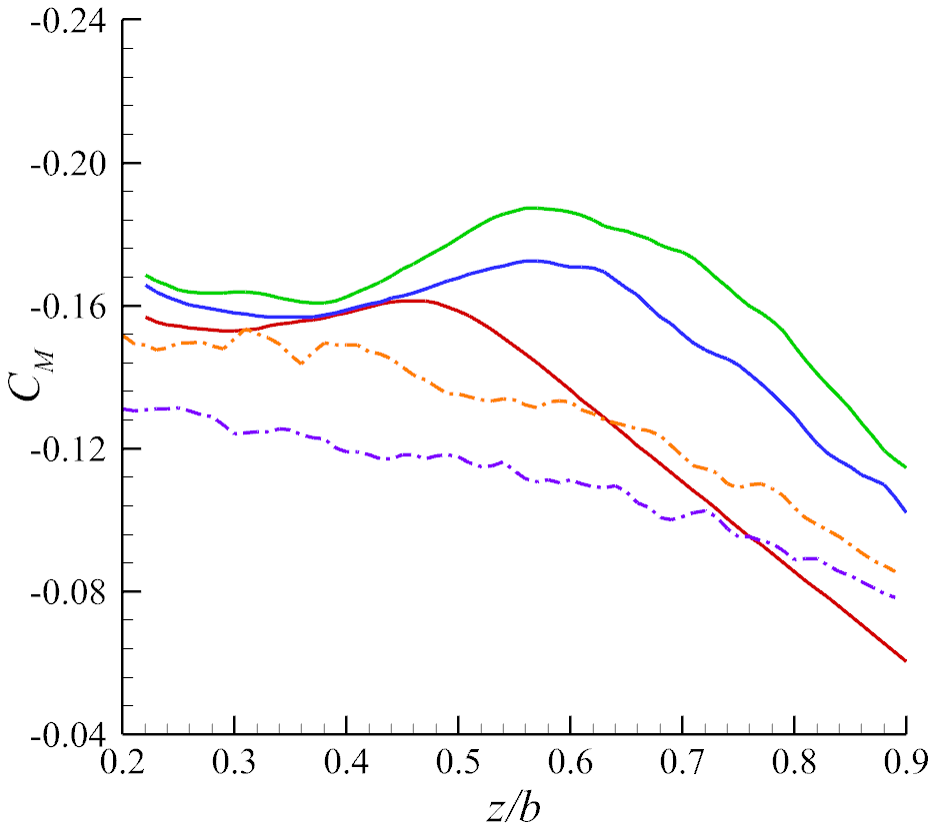}}%
\label{C-z3}
\caption{Sectional forces and pitching moment.}
\label{fig:C-z}
\end{figure}

Figure \ref{fig:C-z} shows the sectional forces and pitching moment for the iced swept wing and the iced straight wing on the fine grid. For the iced swept wing, the lift coefficient distributions simulated by the AMD-IDDES and IDDES methods are similar, but the drag coefficient simulated by the IDDES method is slightly higher than that of the AMD-IDDES method. This is consistent with the findings of a review by Stebbins \cite{stebbinsReviewComputationalMethods2019a}, which notes that drag and moment are more challenging to compute. Furthermore, the lift coefficient predicted by RANS is significantly lower than the experimental value, highlighting the challenges of using RANS for large-scale separated flows. The RANS results for the iced straight wing are not presented in the figure because they did not converge. AMD-IDDES predicts lift coefficients that are more consistent with the experimental results \cite{khodadoustAerodynamicsFiniteWing1995}, while IDDES yields higher lift values, with the discrepancy reaching $3\%$ near the midspan. This is similar to the results computed by Xiao et al. \cite{xiaoEnhancedPredictionThreedimensional2022}, where the lift coefficient predicted by IDDES was overestimated. Furthermore, the drag and moment coefficients predicted by IDDES are higher than those from the AMD-IDDES method, particularly for the moment coefficient, where the difference at the midspan is about $20\%$. By comparing the results of the two cases, it can be observed that the lift coefficient variation across different sections of the iced straight wing is relatively small. In contrast, for the iced swept wing, the wing load on the outer sections significantly decreases. This is a typical stall characteristic of swept wings, indicating that the presence of the sweep angle promotes the stall process.\par
\begin{figure}[!t]
\centering
\subfloat[$z/b=0.17$.]{\includegraphics[width=0.32\columnwidth]{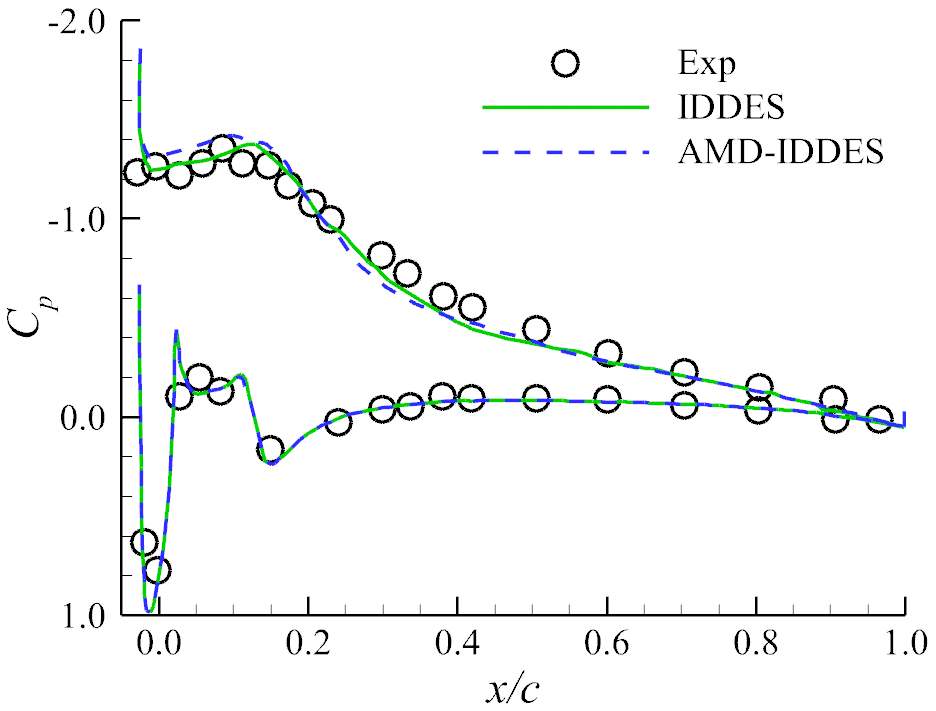}}%
\label{Cp1st}
\hfil
\subfloat[$z/b=0.34$.]{\includegraphics[width=0.32\columnwidth]{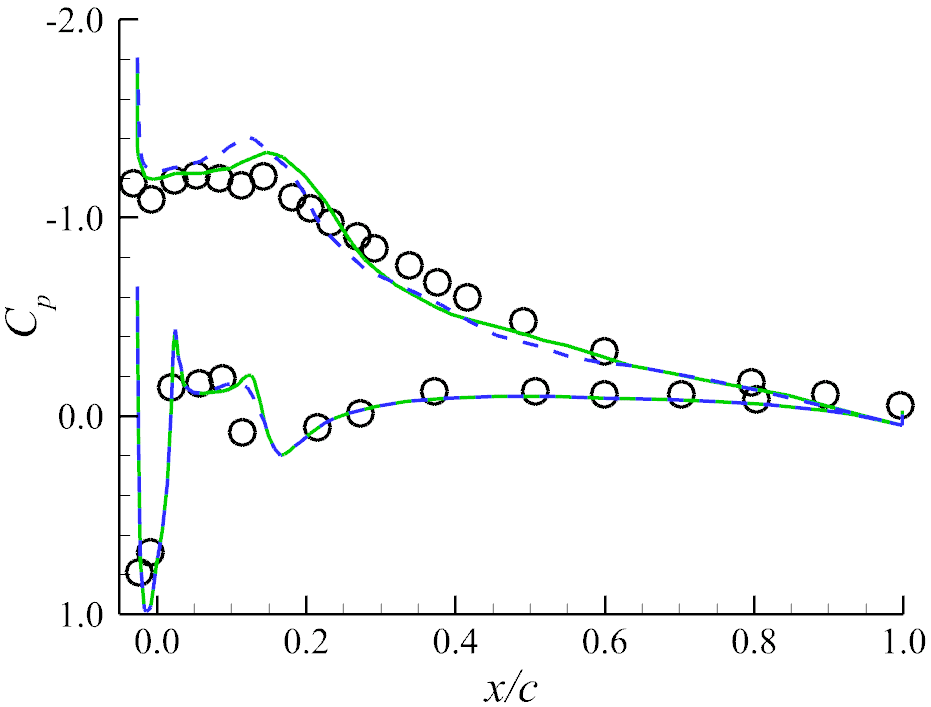}}%
\label{Cp2st}
\subfloat[$z/b=0.50$.]{\includegraphics[width=0.32\columnwidth]{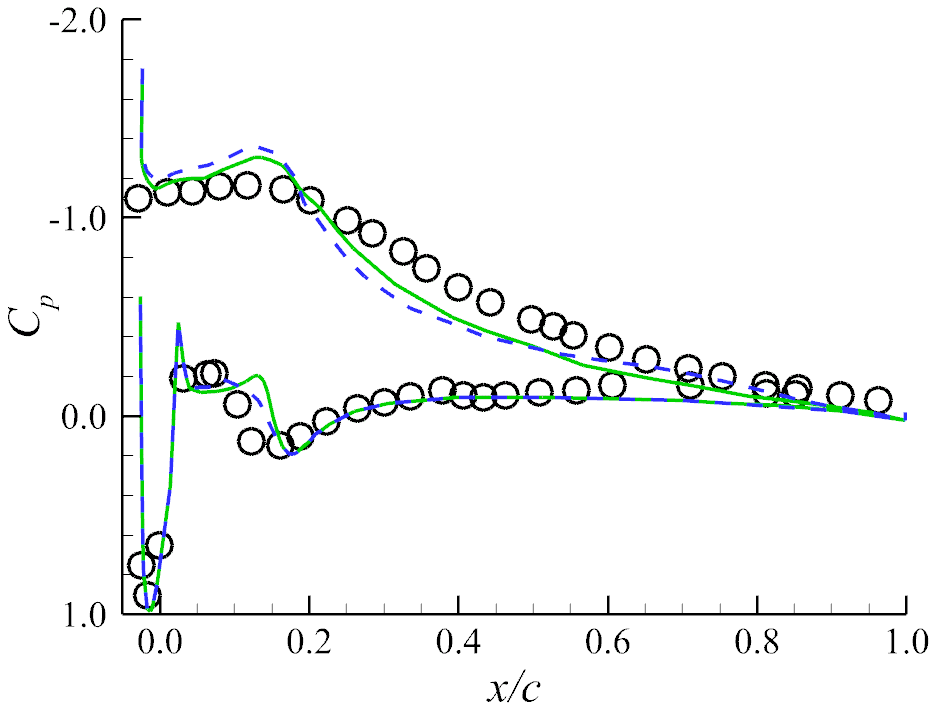}}%
\label{Cp3st} \\
\subfloat[$z/b=0.66$.]{\includegraphics[width=0.32\columnwidth]{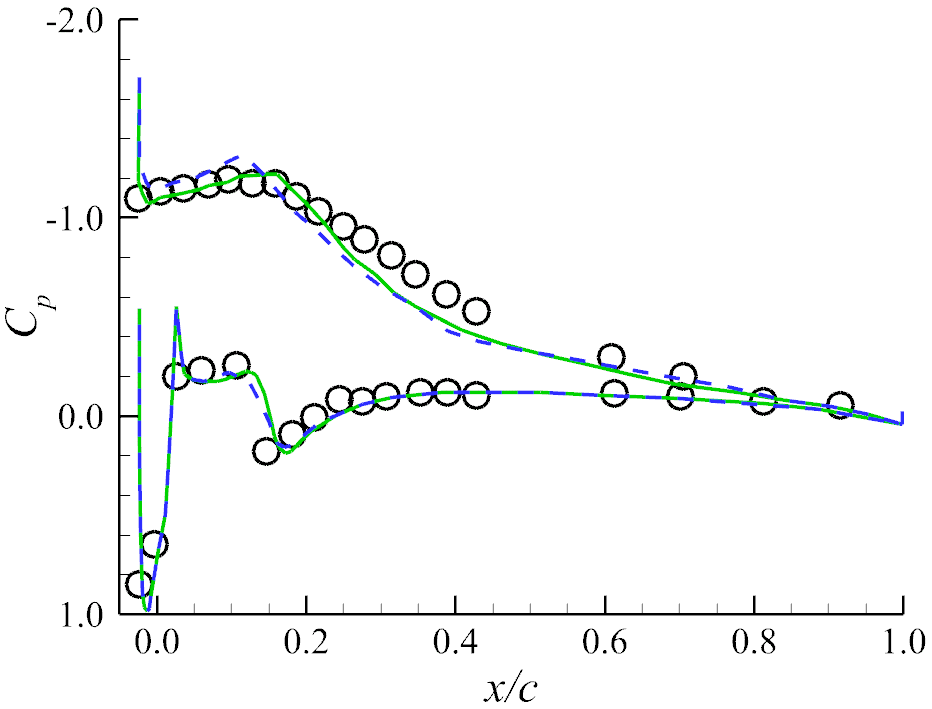}}%
\label{Cp4st}
\subfloat[$z/b=0.85$.]{\includegraphics[width=0.32\columnwidth]{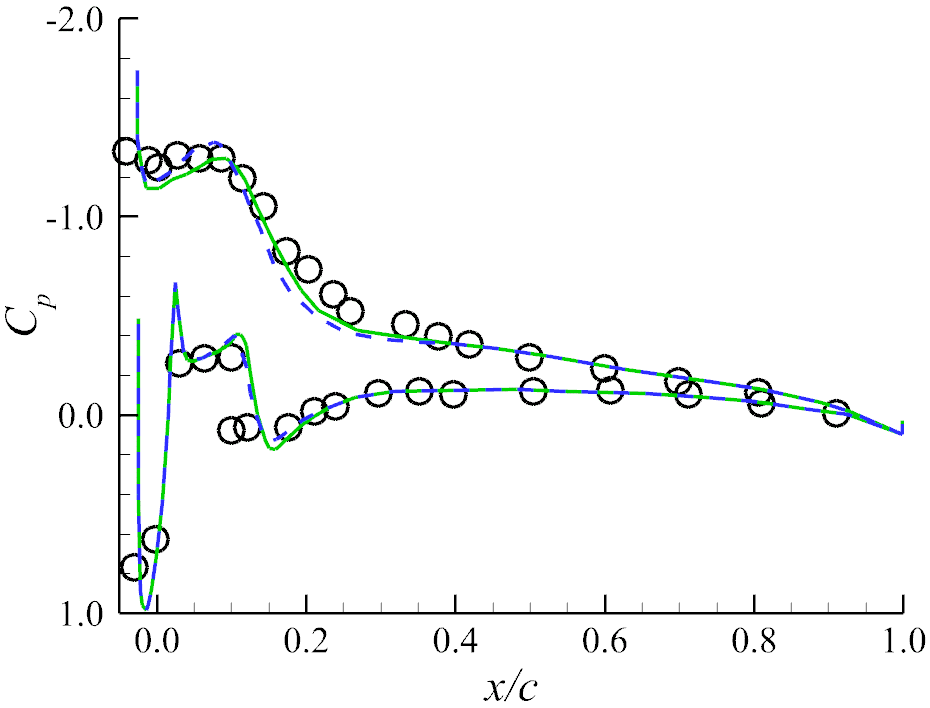}}%
\label{Cp5st}
\caption{The time-averaged surface pressure coefficients from the fine grid for the iced straight wing.}
\label{fig:Cpst}
\end{figure}

Figure \ref{fig:Cpst} compares the time-averaged surface pressure coefficients of the iced straight wing from the fine grid with experimental measurements \cite{khodadoustAerodynamicsFiniteWing1995}. The pressure coefficient distributions predicted by AMD-IDDES on the coarse grid almost coincide with those obtained on the fine grid. Hence, only the results obtained on the fine grid will be analyzed here. Ice accumulation causes flow separation shear layers to form behind the ice, with a surface pressure plateau around the core of the separation bubble. The icing of the leading edge causes the suction peak of the wing's leading edge to be replaced by an almost constant pressure flat top covering about $20\%$ of the chord length. As the flow develops downstream, the pressure gradually recovers, exhibiting characteristics similar to the surface pressure distribution within laminar separation bubbles \cite{leeMechanismsSurfacePressure2015}. The pressure coefficient simulated by IDDES shows a similar distribution to the iced straight wing results obtained by the AMD-IDDES method, a trend that is also consistent with the comparison between IDDES and SLA-IDDES reported by Xiao et al. \cite{xiaoEnhancedPredictionThreedimensional2022}.\par
%

\begin{figure}[!t]
\centering
\subfloat[$z/b=0.27$.]{\includegraphics[width=0.32\columnwidth]{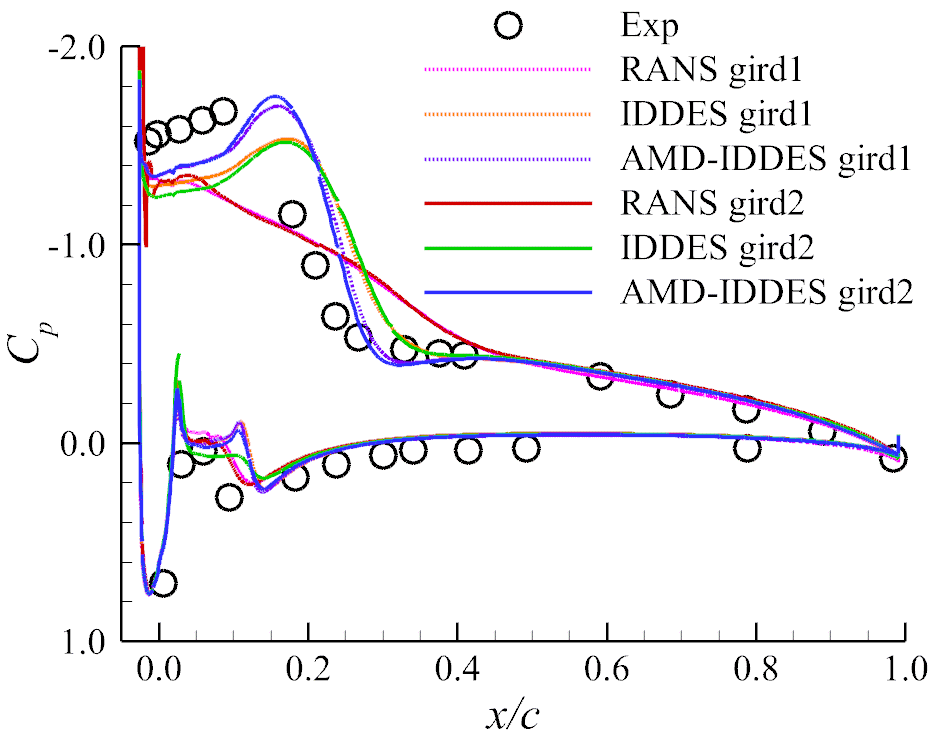}}%
\label{Cp1}
\hfil
\subfloat[$z/b=0.42$.]{\includegraphics[width=0.32\columnwidth]{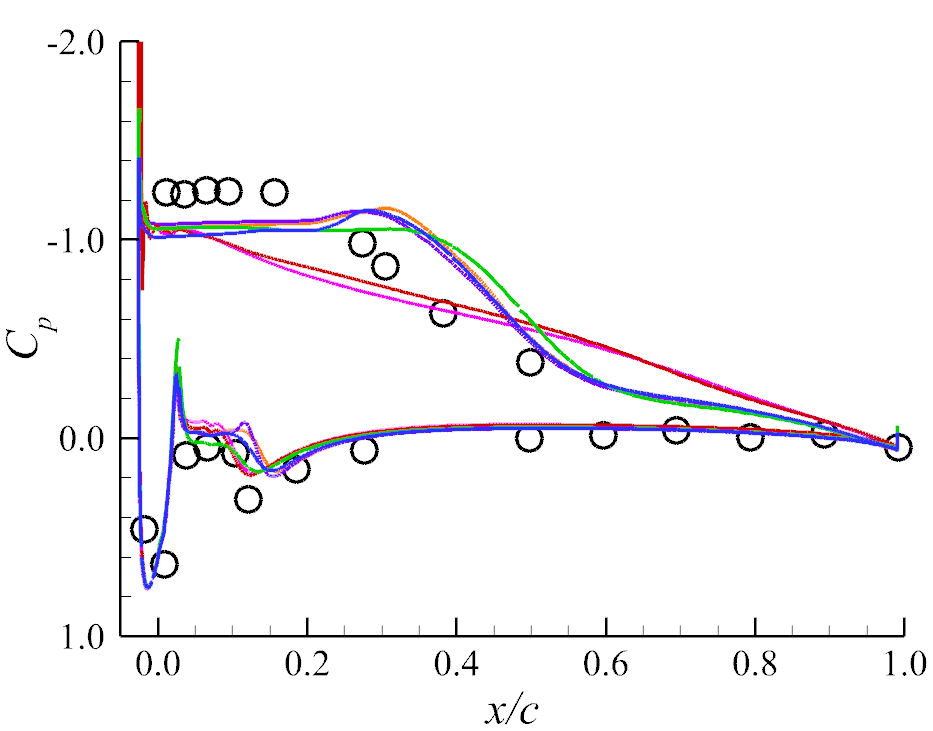}}%
\label{Cp2}
\subfloat[$z/b=0.56$.]{\includegraphics[width=0.32\columnwidth]{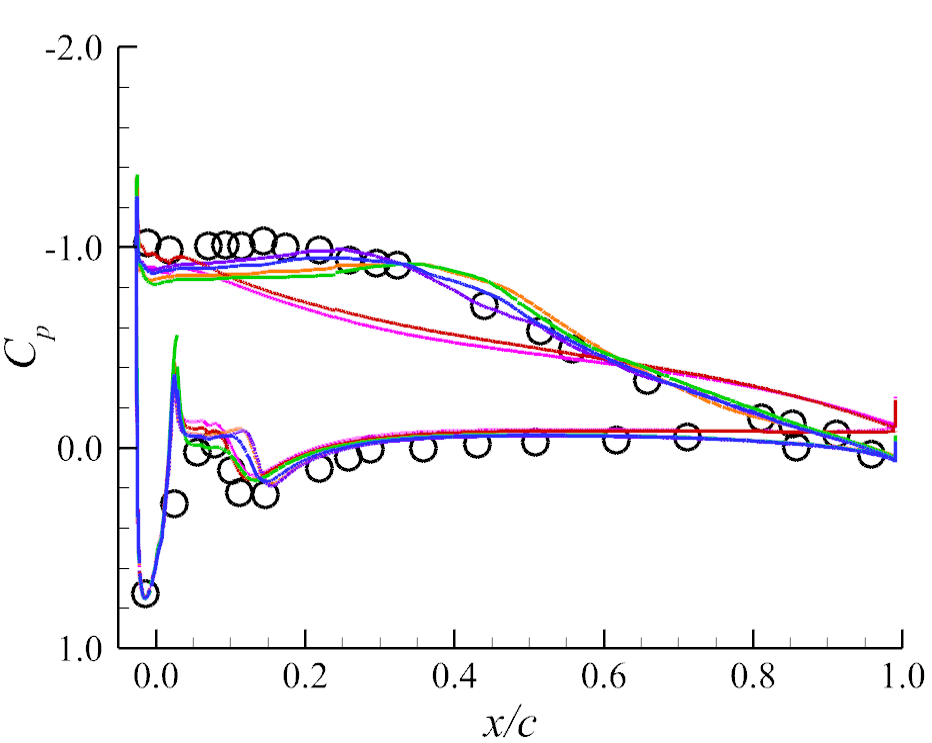}}%
\label{Cp3} \\
\subfloat[$z/b=0.72$.]{\includegraphics[width=0.32\columnwidth]{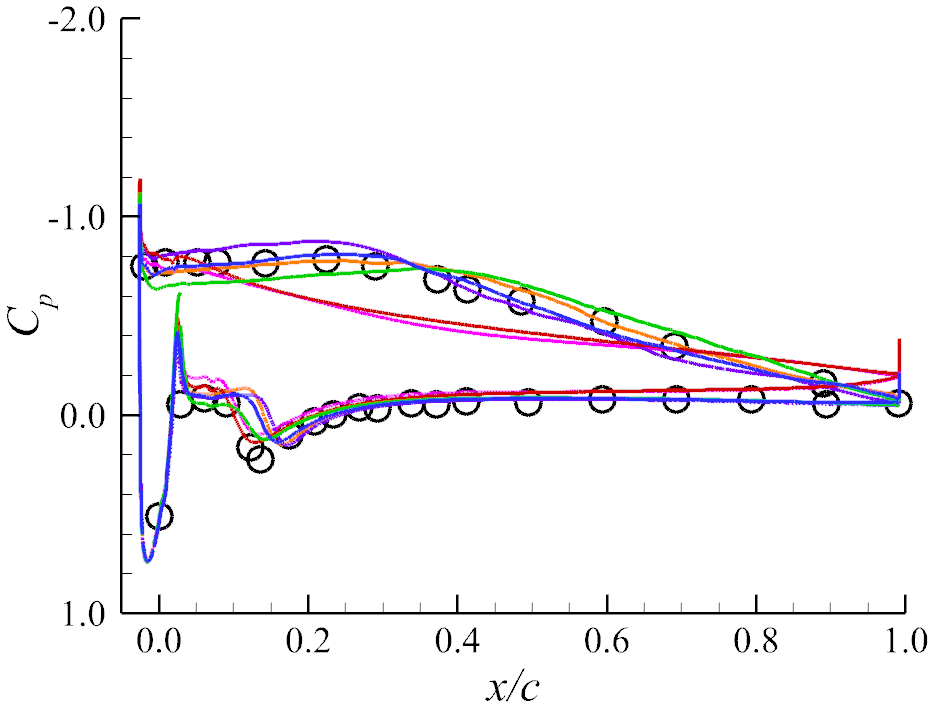}}%
\label{Cp4}
\subfloat[$z/b=0.89$.]{\includegraphics[width=0.32\columnwidth]{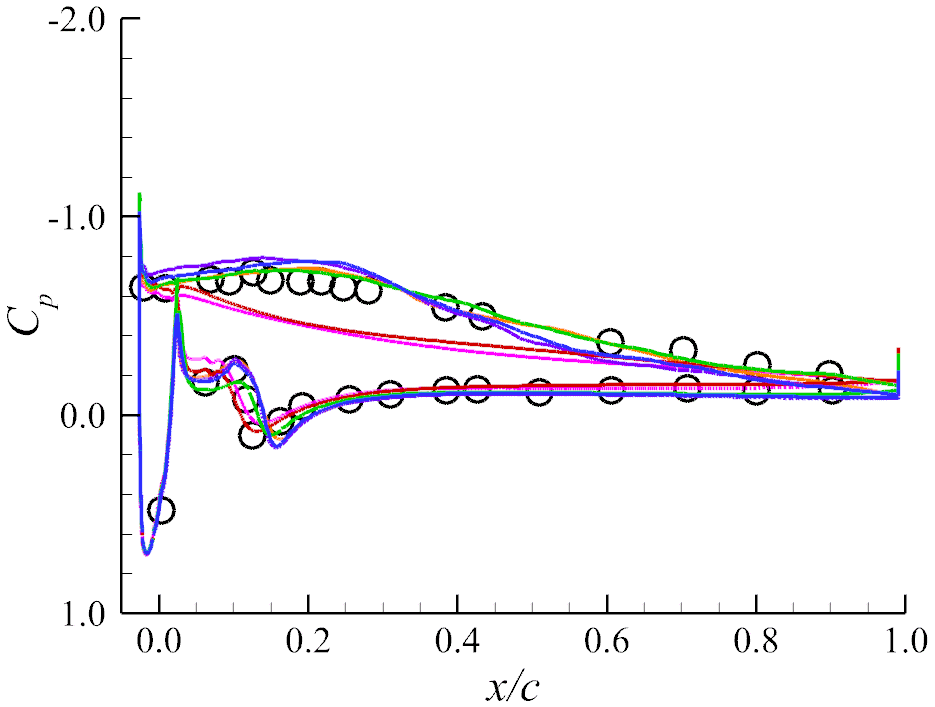}}%
\label{Cp5}
\caption{Time-averaged surface pressure coefficients for the iced swept wing. Grid 1: coarse grid, Grid 2: fine grid.}
\label{fig:Cp}
\end{figure}

%
Figure \ref{fig:Cp} compares the calculated surface pressures of the iced swept wing from the coarse grid and the fine grid with experimental measurements \cite{braggAerodynamicMeasurementsFinite1991}. The comparison between the coarse and fine mesh results indicates that the method exhibits low sensitivity to mesh resolution; therefore, only the fine mesh results are analyzed in the following. The pressure coefficient distribution on the wing root and mid-span regions of the iced swept wing is similar to that of the iced straight wing, with a separation bubble present near the leading edge, followed by a gradual pressure recovery. The difference lies in the spanwise development of the separation region: for the iced swept wing, the separation region extends significantly across the entire upper surface near the wingtip, whereas for the iced straight wing, noticeable differences in the pressure coefficient appear only near the wingtip, with the pressure coefficient distributions at other sections being relatively similar. This indicates that the flow over the iced straight wing mainly exhibits two-dimensional characteristics, whereas the flow over the iced swept wing is significantly influenced by the sweep angle.\par
Comparing the results obtained using the three methods, it can be seen that the pressure coefficients predicted by the RANS method show significant discrepancies with the experimental data. The method struggles to accurately predict the pressure plateau at the leading edge. Both the IDDES and AMD-IDDES methods markedly improve this issue. Compared with IDDES, the absolute value of the pressure coefficient simulated by the AMD-IDDES method is slightly higher in the separation region but slightly lower in the reattachment region. This is similar to the results for the iced straight wing (Figure \ref{fig:Cpst}) and from Xiao et al. \cite{xiaoEnhancedPredictionThreedimensional2022}, where the standard IDDES method slightly stretches and lowers the flat pressure region compared to the two improved IDDES methods. Due to the “gray area” issue, the standard IDDES method delays the onset of instability, causing the recirculation region to be elongated.\par
\section{Flow separation characteristics and vortex dynamics}\label{sec:Vortex Structures and Flow Characteristics}
\subsection{Reliability of the AMD-IDDES method}
\begin{figure}[!t]
\centering
\subfloat[IDDES]{\includegraphics[width=0.538\columnwidth]{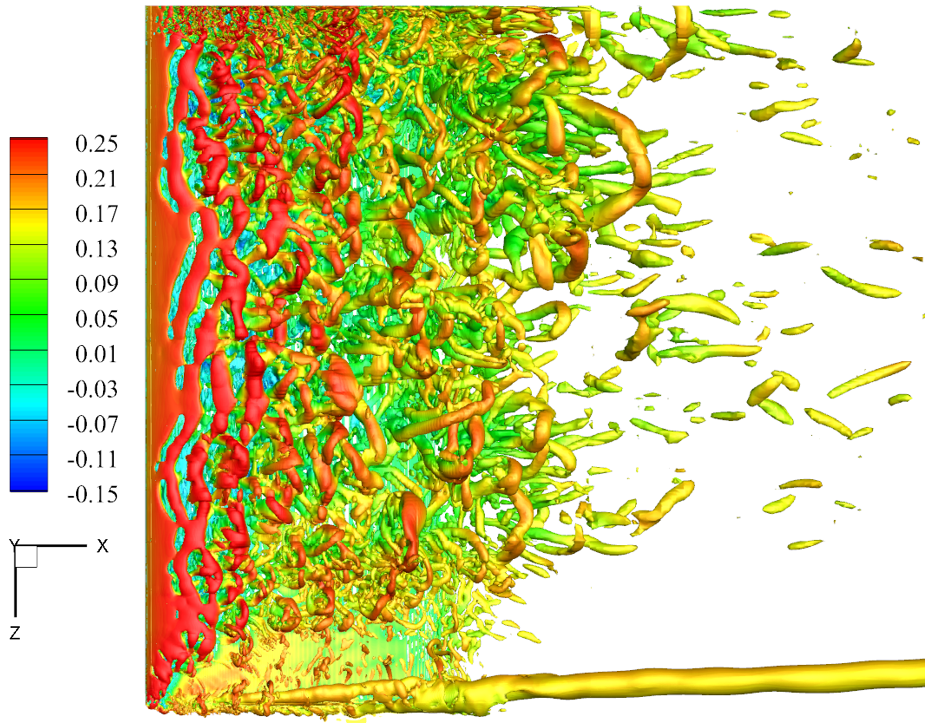}%
\label{Q1st}}
\hfil
\subfloat[AMD-IDDES]{\includegraphics[width=0.462\columnwidth]{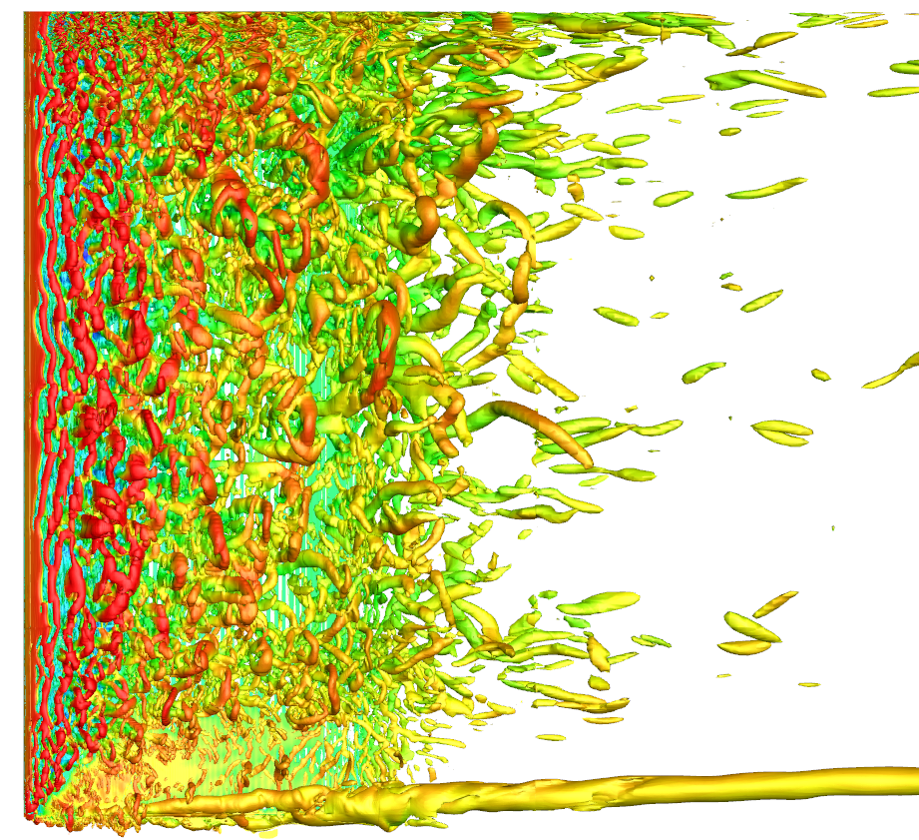}}%
\label{Q2st}
\caption{The instantaneous isosurface $Q(c/U_\infty)=1$ of the iced straight wing simulated by AMD-IDDES, colored by the streamwise velocity $u/U_\infty$.}
\label{fig:Qst}
\end{figure}

\begin{figure}[!t]
\centering
\subfloat[IDDES]{\includegraphics[width=0.525\columnwidth]{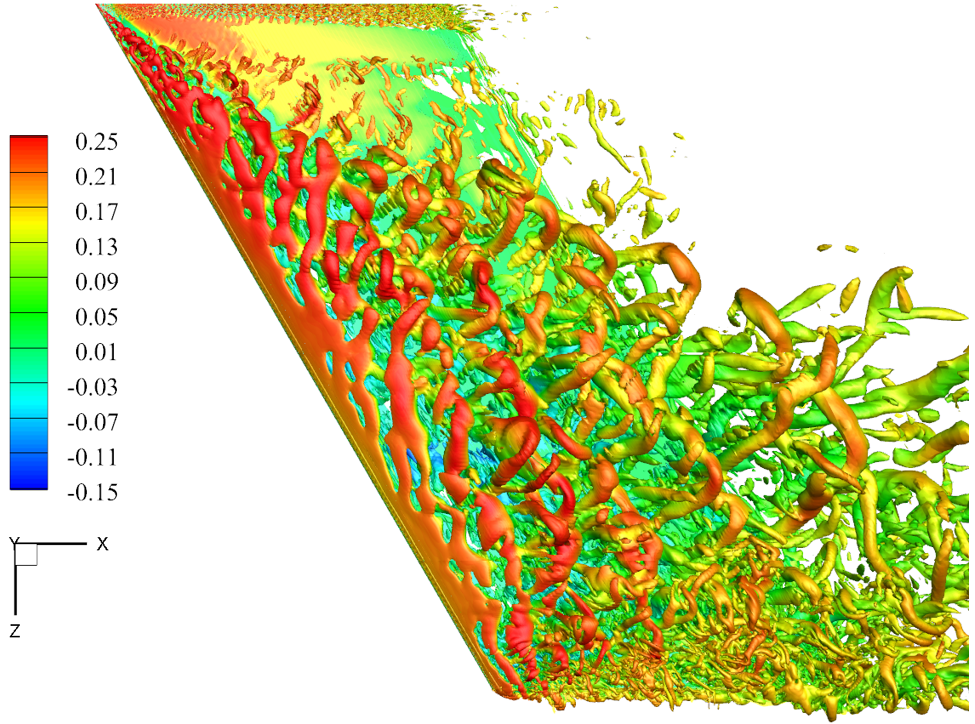}%
\label{Q1}}
\hfil
\subfloat[AMD-IDDES]{\includegraphics[width=0.475\columnwidth]{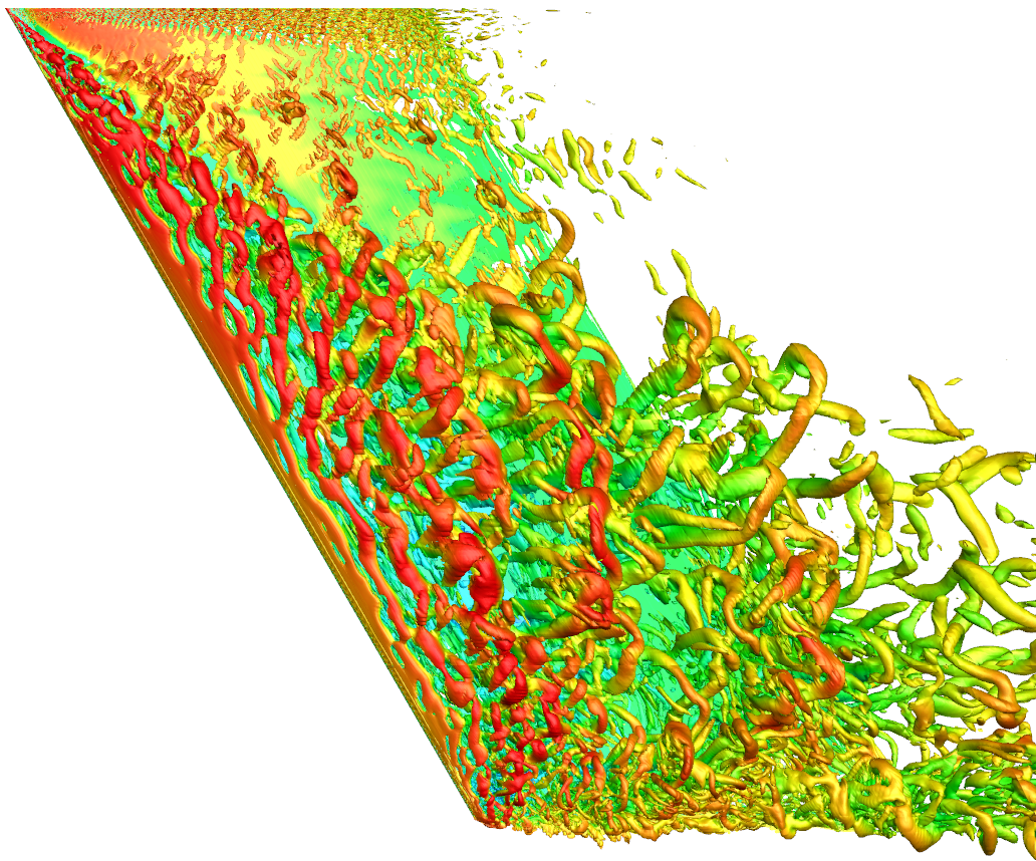}}%
\label{Q2}
\caption{The instantaneous isosurface $Q(c/U_\infty)=1$ of the iced swept wing simulated by AMD-IDDES, colored by the streamwise velocity $u/U_\infty$.}
\label{fig:Q}
\end{figure}

Figures \ref{fig:Qst} and \ref{fig:Q} compare the Q-criterion isosurface structures of the IDDES and AMD-IDDES methods for the iced swept wing and iced straight wing. It can be observed that the AMD-IDDES method predicts an earlier transition process after the leading-edge icing, while the IDDES method results in larger vortex structures downstream. Deck et al. \cite{deckUnsteadinessAxisymmetricSeparatingreattaching2007} referred to the large-scale structures as “overcoherent” structures and indicated that they result in strong fluctuations near reattachment, which is consistent with the extended regions of strong fluctuations yielded by IDDES. Additionally, the AMD-IDDES method provides a clearer simulation of the premature separation phenomenon caused by the downwash at the wingtip, drawing from the flow behavior in the adjacent regions. The results show similarities to previous studies comparing IDDES and SLA-IDDES for iced straight wings \cite{xiaoEnhancedPredictionThreedimensional2022}. \par
\begin{figure}[!t]
\centering
\subfloat[IDDES]{\includegraphics[width=1.0\columnwidth]{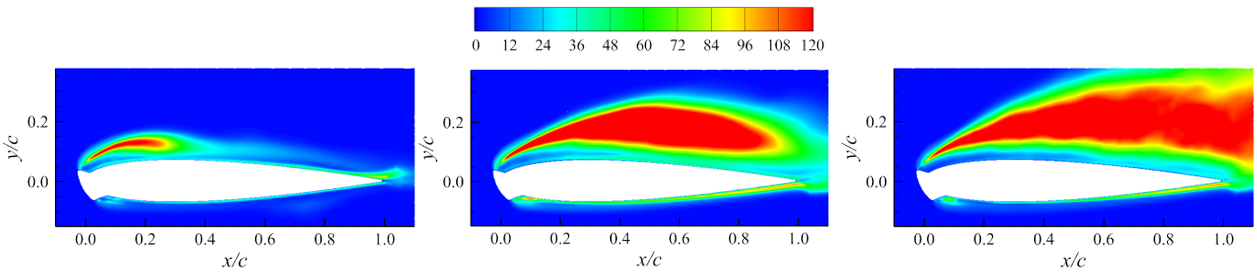}}%
\label{miu2} \\
\subfloat[AMD-IDDES]{\includegraphics[width=1.0\columnwidth]{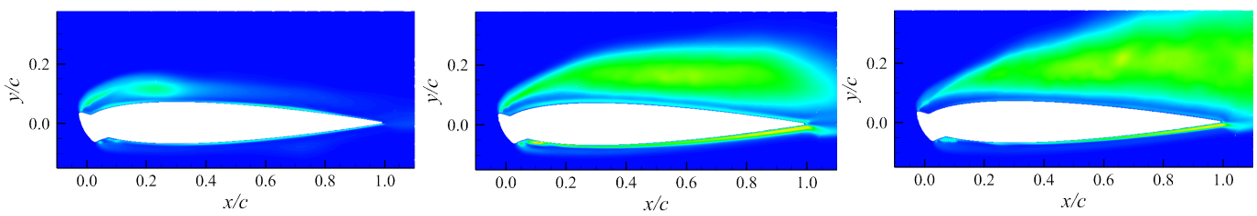}}%
\label{miu3} 
\caption{The time-averaged eddy viscosity $\mu _t/\mu _\infty$ at different spanwise locations. Left: $z/b=0.27$; middle: $z/b=0.56$; right: $z/b=0.89$ for the iced swept wing.}
\label{fig:miu}
\end{figure}

Figure \ref{fig:miu} shows the time-averaged eddy viscosity $\mu _t/\mu _\infty$ at different spanwise locations for the iced swept wing. In the shear layer region, the eddy viscosity predicted by the IDDES simulation is significantly higher than that of the AMD-IDDES method. This is consistent with the results from §\ref{sec:Calibration and Test}, where the IDDES method suffers from excessive dissipation due to its length scale definition, leading to a delayed onset of the K--H instability. As a result, it struggles to capture the subtle fluctuations during shear layer development. In contrast, the AMD-IDDES method mitigates this issue by improving the definition of the length scale. Therefore, the AMD-IDDES method offers clear advantages in industrial applications, particularly in noise prediction, by providing more accurate representations of flow dynamics and turbulence fluctuations. \par

%
Although the AMD-IDDES method shows certain advantages over the IDDES method in resolving flow fluctuations and vortex structures, the difference between the two methods in simulating the pressure coefficient (Figure \ref{fig:Cpst} and Figure \ref{fig:Cp}) is not significant compared to the experimental results. Hence, the subsequent discussion shifts from method comparison to analyzing the flow fields obtained by AMD-IDDES for iced straight and swept wings, in order to examine how icing influences the flow characteristics of swept wings. \par
\subsection{Primary flow characteristics induced by the sweep angle}
\begin{figure}[!t]
\centering
\subfloat[Iced straight wing.]{\includegraphics[width=1.0\columnwidth]{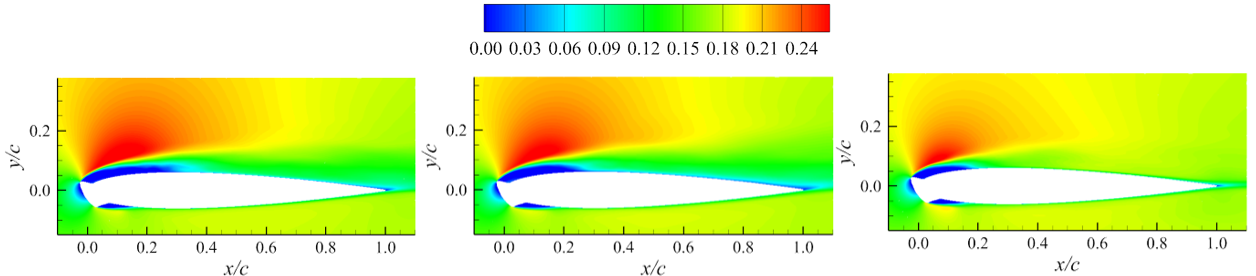}}%
\label{uavgx1}
\hfil \\
\subfloat[Iced swept wing.]{\includegraphics[width=1.0\columnwidth]{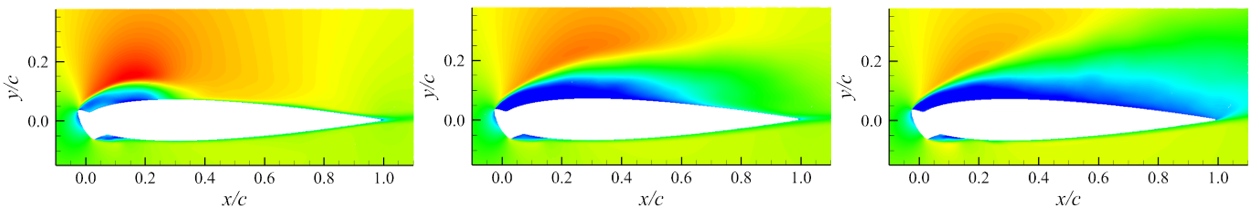}}%
\label{uavgx2} \\
\caption{The time-averaged streamwise velocity at different spanwise locations for the iced straight wing and iced swept wing. Left: $z/b=0.17$; middle: $z/b=0.50$; right: $z/b=0.85$ for the iced straight wing. Left: $z/b=0.27$; middle: $z/b=0.56$; right: $z/b=0.89$ for the iced swept wing.}
\label{fig:uavgx}
\end{figure}

Figure \ref{fig:uavgx} shows the time-averaged streamwise velocity $U_x/U_{\infty}$ fields for the iced straight wing overlaid by the stream locations for $z/b=0.17$, 0.50, and 0.85, and for the iced swept wing overlaid by the stream locations for $z/b=0.27$, 0.56, and 0.89. The figure provides an overall picture of the variation of the separation bubble along the spanwise direction. The distribution trend of the separation bubble is basically consistent with the pressure plateau distribution observed in the pressure coefficient. \par
\begin{figure}[hbt!]
  \centering
  \includegraphics[width=0.80\columnwidth]{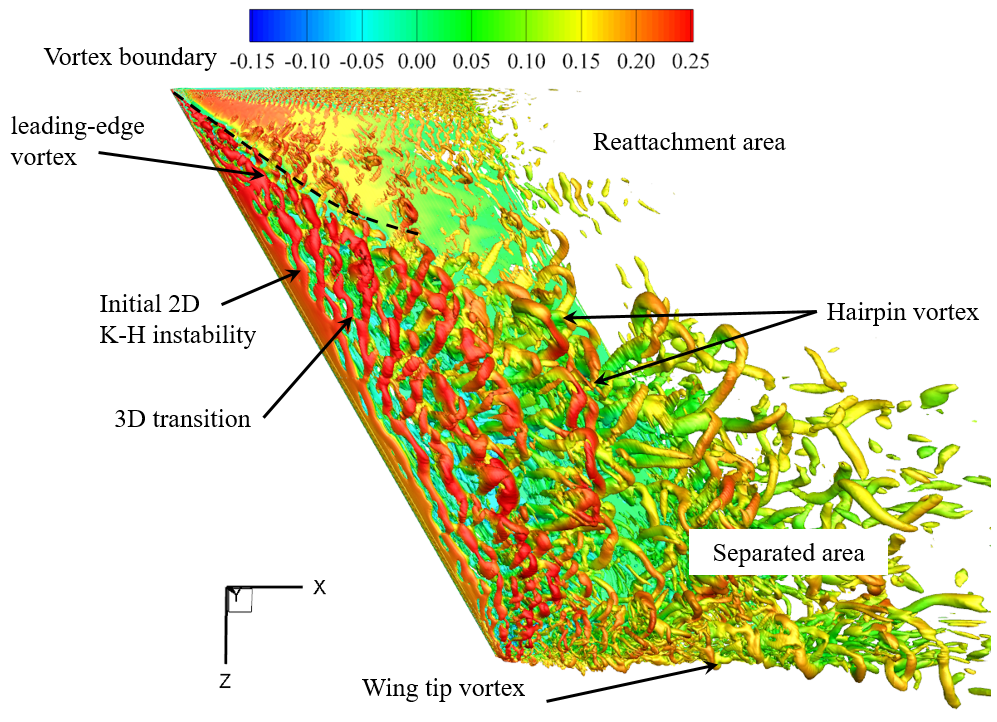}\hfill
  \caption{The instantaneous isosurface $Q(c/U_\infty)=1$ of the iced swept wing simulated by AMD-IDDES and colored by the streamwise velocity $u/U_\infty$.} 
  \label{fig:Q_stucture}
\end{figure}

%
Figure \ref{fig:Q_stucture} shows the flow structures identified from the Q iso-surface for the iced swept wing. Flow over the ice shape generates a strong adverse pressure gradient, causing flow separation and forming a leading-edge separation bubble. Inside the separation bubble, a large recirculation region develops. The flow starts to reattach downstream of the bubble, forming a reattachment line. Near the reattachment line, the flow velocity increases and the pressure decreases, creating a low-pressure zone. The suction effect from this low-pressure zone drives a reverse flow from the trailing edge toward the leading edge, forming a vortex structure. This behavior is similar for straight and swept wings. However, since straight wings lack a spanwise flow, the overall flow structure remains similar to a two-dimensional flow pattern. \par

On a swept wing, the sweep angle induces a significant spanwise flow. The spanwise flow redirects the disturbances within the separation bubble along the spanwise direction, altering the structure of the recirculation region. This transformation leads to the evolution of the separation region from a localized two-dimensional pattern to a three-dimensional structure, which triggers three-dimensional instabilities and induces vortex formation. Under the combined effect of shear and the pressure gradient, the spanwise flow near the leading-edge separation region induces the formation of a vortex called the ``leading-edge vortex'' extending along the spanwise direction. \par
\begin{figure}[hbt!]
  \centering
  \includegraphics[width=0.6\columnwidth]{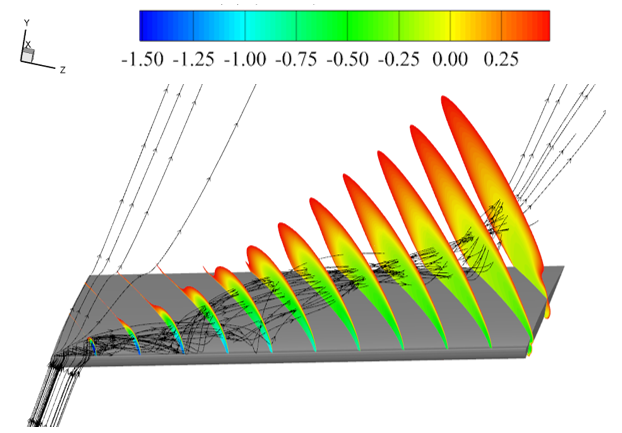}\hfill
  \caption{Spatial distributions of the total pressure coefficient $C_{pt}$ and streamlines for the iced swept wing.} 
  \label{fig:Cpt1}
\end{figure}

Around the vortex core, there are pronounced variations in the velocity and pressure that form a transition zone between the vortex core and the boundary layer, which defines the “vortex boundary" \cite{khodadoustAerodynamicsFiniteWing1995}. As a result, the overall flow is naturally divided by the vortex boundary into a reattachment region and a separation region. At the leading edge of the separation region, the vortex evolution is evident, transitioning from the initial two-dimensional Kelvin–Helmholtz (K--H) instability into a three-dimensional structure. As these vortices are convected downstream, they grow in size and develop into more complex three-dimensional structures. Figure \ref{fig:Cpt1} shows the spatial distribution of the total pressure coefficient, $C_{pt}$, for the iced swept wing. Its distribution illustrates the development of the leading-edge vortex along the spanwise direction. The overall trend of the streamlines is consistent with the experimental results \cite{khodadoustAerodynamicsFiniteWing1995}. \par
%
\begin{figure}[!t]
\centering
\subfloat[Iced straight wing.]
{\includegraphics[width=1.0\columnwidth]{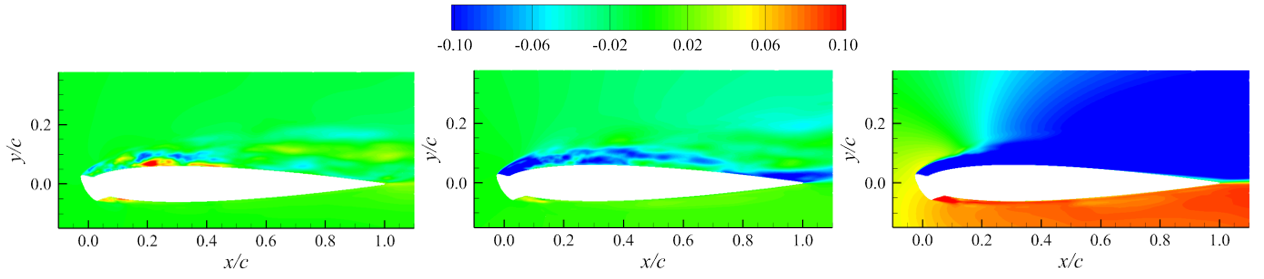}}%
\label{uz1} 
\subfloat[Iced swept wing.]
{\includegraphics[width=1.0\columnwidth]{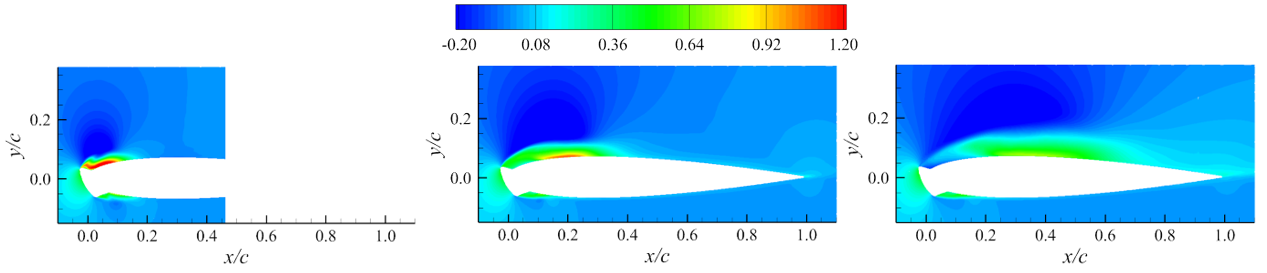}}%
\label{uz2} 
\caption{The time-averaged spanwise velocity $U_z/U_{\infty}$ at different spanwise locations. Left: $z/b=0.17$; middle: $z/b=0.50$; right: $z/b=0.85$ for the iced straight wing. Left: $z/b=0.10$; middle: $z/b=0.27$; right: $z/b=0.42$ for the iced swept wing.}
\label{fig:uz}
\end{figure}

Within the leading-edge vortex, the spanwise velocity is noticeably high. Figure \ref{fig:uz}(b) shows the time-averaged spanwise velocity for the iced swept wing at different spanwise locations. The cross-sectional locations are shown in Figure \ref{fig:model}. In addition to the positions previously used for the iced swept wing, an extra cross-section at $z/b=0.1$ is also extracted to illustrate the flow distribution near the wing root. The extraction method for this section is similar to that of the other sections. In the wing-root region, the spanwise velocity is particularly strong, reaching nearly twice the free-stream velocity, consistent with the findings of Kwon et al. \cite{kwonNumericalStudyEffects1991}. As the flow progresses toward the wingtip, the spanwise velocity gradually decreases. This trend is attributed to the significant three-dimensional flow interactions near the wing root, where a strong spanwise pressure gradient accelerates the flow in the spanwise direction. Moving toward the wingtip, the pressure gradient weakens, reducing the spanwise flow momentum. In contrast, Figure \ref{fig:uz}(a) shows that the overall spanwise velocity on the iced straight wing is relatively low. Only the wingtip region is significantly influenced by the wingtip vortex, which induces a negative spanwise velocity on the upper surface and a positive velocity on the lower surface. Due to the effect of the wingtip vortex, a region of negative and relatively small spanwise velocity also appears on the upper surface of the mid-span area.\par
%

\begin{figure}[!t]
\centering
\subfloat[Iced straight wing.]{\includegraphics[width=1.0\columnwidth]{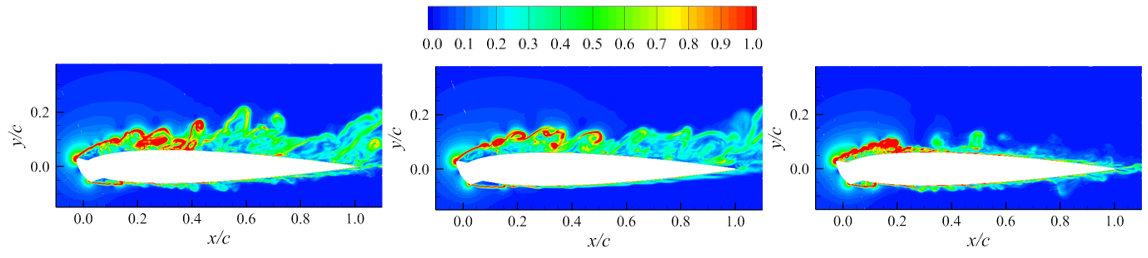}}%
\label{dr1} \\
\subfloat[Iced swept wing.]{\includegraphics[width=1.0\columnwidth]{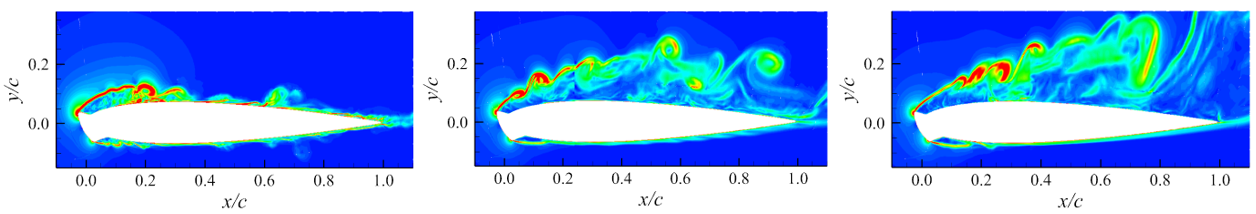}}%
\label{dr2} 
\caption{Instantaneous $\nabla \rho$ at different spanwise locations. Left: $z/b=0.17$; middle: $z/b=0.50$; right: $z/b=0.85$ for the iced straight wing. Left: $z/b=0.27$; middle: $z/b=0.56$; right: $z/b=0.89$ for the iced swept wing.}
\label{fig:grad_rho}
\end{figure}

\begin{figure}[!t]
\centering
\subfloat[Iced straight wing.]{\includegraphics[width=0.387\columnwidth]{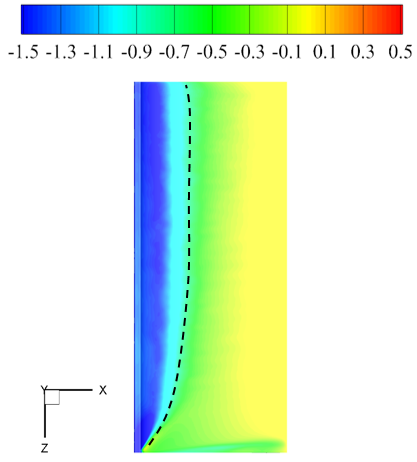}}%
\label{flow_model1}
\hfil 
\subfloat[Iced swept wing.]{\includegraphics[width=0.593\columnwidth]{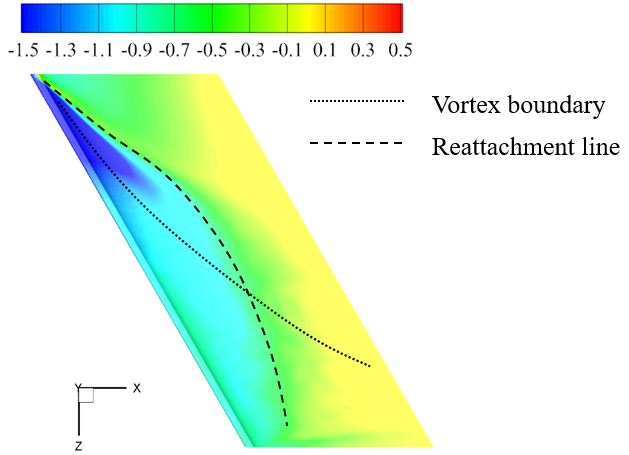}}%
\label{flow_model2} \\
\caption{Vortex boundary and reattachment line. (Adapted from Khodadoust et al. \cite{khodadoustAerodynamicsFiniteWing1995}.)}
\label{fig:flow_model}
\end{figure}

%
%
Figure \ref{fig:grad_rho} shows the instantaneous $\nabla \rho$ at different spanwise locations. The development trend of the separation region shown in this figure is consistent with the trends presented in Figure \ref{fig:Cpst}, Figure \ref{fig:Cp}, and Figure \ref{fig:uavgx}. Additionally, the reattachment locations can be obtained using the time-averaged velocity distribution (Figure \ref{fig:uavgx}), as shown in Table \ref{tbl:reattachment}. Analysis of these results indicates that the separation region on the straight wing is largest in the mid-span region, slightly smaller near the root, and smallest near the tip. A more detailed theoretical explanation of this spanwise variation can be found in Xiao et al. \cite{xiaoEnhancedPredictionThreedimensional2022}. For the swept wing, the separation region increases progressively in the spanwise direction. By analyzing the reattachment region and the leading-edge vortex, flow schematics for the iced straight wing and the iced swept wing were obtained (Figure \ref{fig:flow_model}).\par
%
\begin{table}[hbt!]
  \centering
qept  \caption{Reattachment locations.}
  \label{tbl:reattachment}
  \begin{tabular}{ c c c c c c }
    \hline
    \hline
    \mc{$L/c$}    & \mc{Slice 1} & \mc{Slice 2} & \mc{Slice 3} & \mc{Slice 4} & \mc{Slice 5}   \\
    \hline
    \small Iced straight wing & $0.26$  & $0.27$  & $0.27$ & $0.27$  & $0.18$  \\
    \small Iced swept wing & $0.24$  & $0.45$ & $0.63$ & $0.77$  & $0.85$     \\
    \hline
    \hline
  \end{tabular}
\end{table}\par
\subsection{Influence of the wing root and wingtip on the primary flow}
\begin{figure}[!t]
\centering
\subfloat[Wing root.]{\includegraphics[width=0.49\columnwidth]{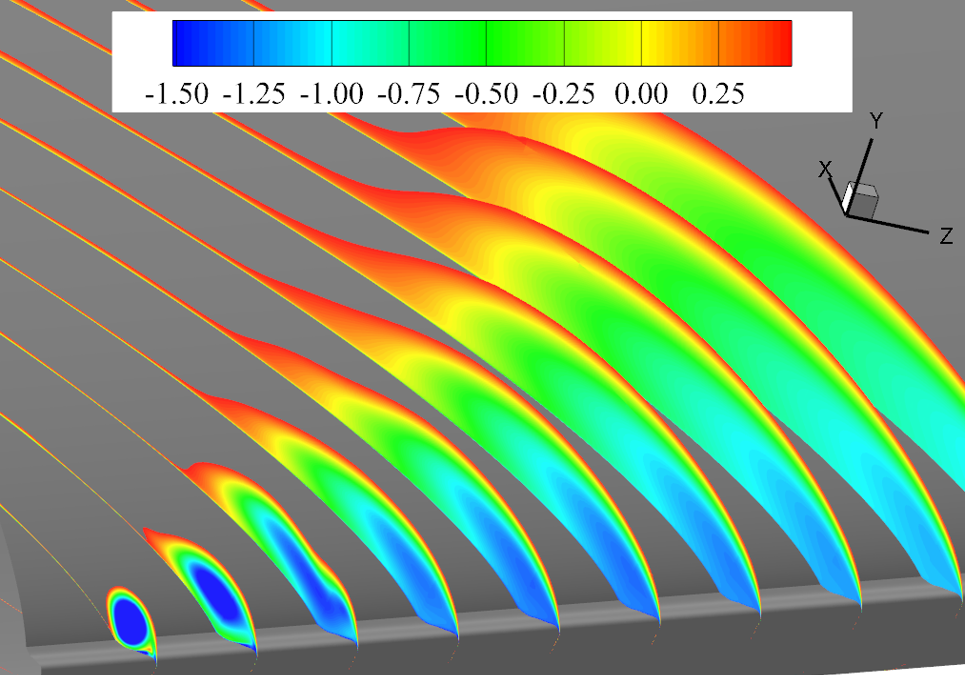}}%
\label{Cpt_1}
\subfloat[Wing tip.]{\includegraphics[width=0.45\columnwidth]{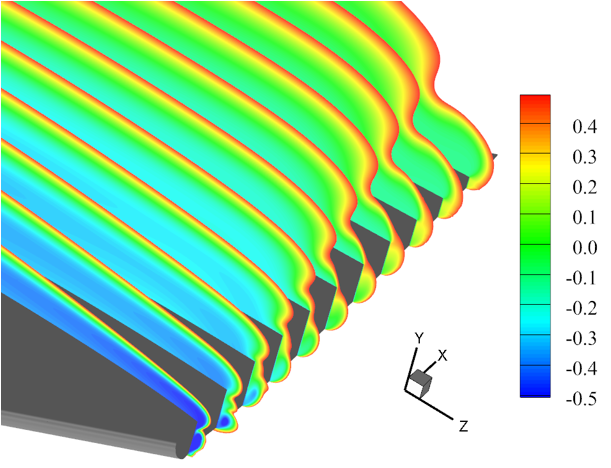}}%
\label{Cpt_2} \\
\caption{Local spatial distributions of the total pressure coefficient $C_{pt}$ for the iced swept wing.}
\label{fig:Cpt_local}
\end{figure}

The effects of the sweep angle on the flow characteristics of both the iced straight and swept wings have now been analyzed. In addition, both wing root and wing tip effects have been clearly observed in the previous sections. Figure \ref{fig:Cpt_local} illustrates the $C_{pt}$ distribution near the wing root and wing tip for the iced swept wing. For a more detailed analysis of the wing tip and wing root structure of the iced straight wing, refer to Xiao et al. \cite{xiaoEnhancedPredictionThreedimensional2022}. \par
%
For the iced straight wing, the end-wall effect significantly influences the separation bubble in the wing root region.  This is due to the vortical structures (such as the horseshoe vortex in Figure \ref{fig:Qst}) induced by the end-wall effect, which distort the flow field and cause the flow to reattach earlier in the wing root region, thus reducing the size of the separation bubble. In contrast, for the iced swept wing, the significant spanwise flow induced by the sweep angle has a noticeable impact on the flow in the wing root region.  The spanwise flow transports low-energy fluid along the span, preventing the accumulation and amplification of potential separation vortices at the wing root, thus mitigating the wing root effect.  Furthermore, compared to the development of the leading-edge vortex along the spanwise direction [shown in Figure \ref{fig:Cpt_local}(a)], the wing root effect during the initial development of the leading-edge vortex does not significantly affect the flow structure.  Therefore, the wing root effect is less pronounced in the iced swept wing than in the iced straight wing.\par
%
Due to the pressure difference between the upper and lower surfaces of the wing, the flow on the lower surface near the wing tip rolls upward, forming a tip vortex. For the straight wing, the tip vortex dominates the flow in the wing tip region, where the vortex structures generated by the separation shear layer due to ice formation are relatively weak. This is because the tip vortex induces a downwash on the main flow, decelerating the flow near the wing tip, increasing the local pressure and reducing the pressure gradient. The dominant flow structure for the straight wing is the shear layer vortex developing along the streamwise direction. The reduced pressure gradient caused by the tip vortex suppresses the further development of the separated shear layer, making it more difficult for the separated region to generate coherent vortex structures and inhibiting the formation of streamwise vortices.\par
For the swept wing, the dominant flow structure is the leading-edge vortex with a significant spanwise flow. The presence of this vortex results in a large spanwise velocity near the wing tip, which reduces the suction peak (as shown in Figure \ref{fig:Cp}) and reduces the pressure difference between the upper and lower surfaces, making the formation of the tip vortex more difficult. However, since a pressure difference persists, a tip vortex still forms, but it is less distinct in the Q-criterion structures than for the straight wing. Nevertheless, the flow structure induced by the tip vortex can still be observed in the $C_{pt}$ distribution [Figure \ref{fig:Cpt_local}(b)]. Furthermore, the tip vortex can weaken the formation of the leading-edge vortex and reduce its strength and stability, thereby limiting its further spanwise development beyond the wing tip region.\par
\begin{figure}[!t]
\centering
\subfloat[Iced straight wing.]{\includegraphics[width=1.0\columnwidth]{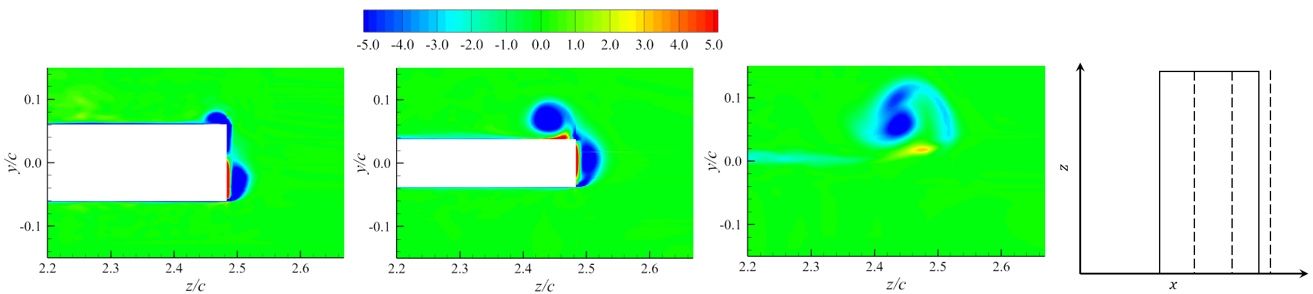}}%
\label{vortx_straight}
\hfil \\
\subfloat[Iced swept wing.]{\includegraphics[width=1.0\columnwidth]{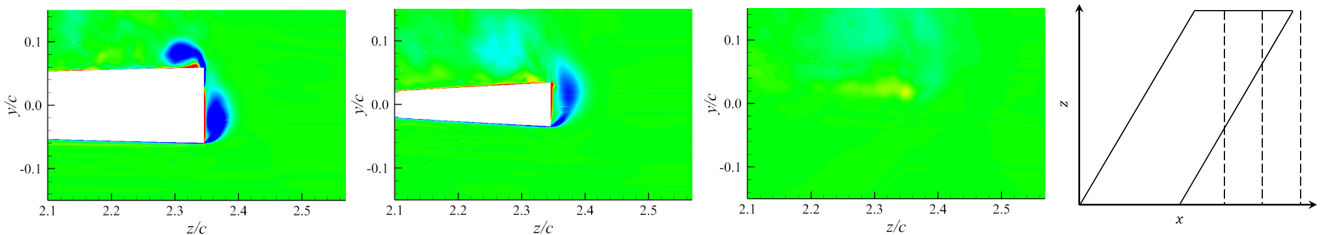}}%
\label{vortx_swept} 
\caption{The time-averaged $x$-vorticity distributions for the iced swept and straight wings.}
\label{fig:vortx}
\end{figure}

Figure \ref{fig:vortx} shows the development of the time-averaged $x$-vorticity distribution along the streamwise direction for both wing configurations. The development of the tip vortex shown is fully consistent with the above analysis. Initially, the pressure difference between the upper and lower surfaces is similar for both cases, resulting in comparable tip vortices rolling up from the lower surface. However, as the flow develops downstream, the tip vortex of the straight wing gradually strengthens and increases in size, whereas the tip vortex of the swept wing weakens due to the influence of the leading-edge vortex. \par

\section{Unsteady characteristics of dominant flow structures}\label{sec:Pressure Fluctuation PSD Analysis}
\subsection{Feature frequency analysis}
PSD analyses of the pressure fluctuations for the iced swept wing allow for a deeper investigation into the flow characteristics. All analyses regarding pressure fluctuations are based on the results from the fine grids. The PSD approach was employed to analyze how unsteadiness is distributed in the frequency domain. The nondimensional time step used in the simulation is $\Delta t U_{\infty}/c = 0.01$. To obtain a smoother PSD estimate, a segment-averaging technique was employed. For each calculation, two adjacent segments with a combined length of $2n$ and $50\%$ overlap were selected. A periodogram was then computed using the Fast Fourier Transform (FFT), and the resulting one-sided spectra were averaged across all windows to reduce variance. This method provides a good balance between frequency resolution and statistical reliability in the PSD estimation.\par
\begin{figure}[hbt!]
  \centering
  \includegraphics[width=0.55\columnwidth]{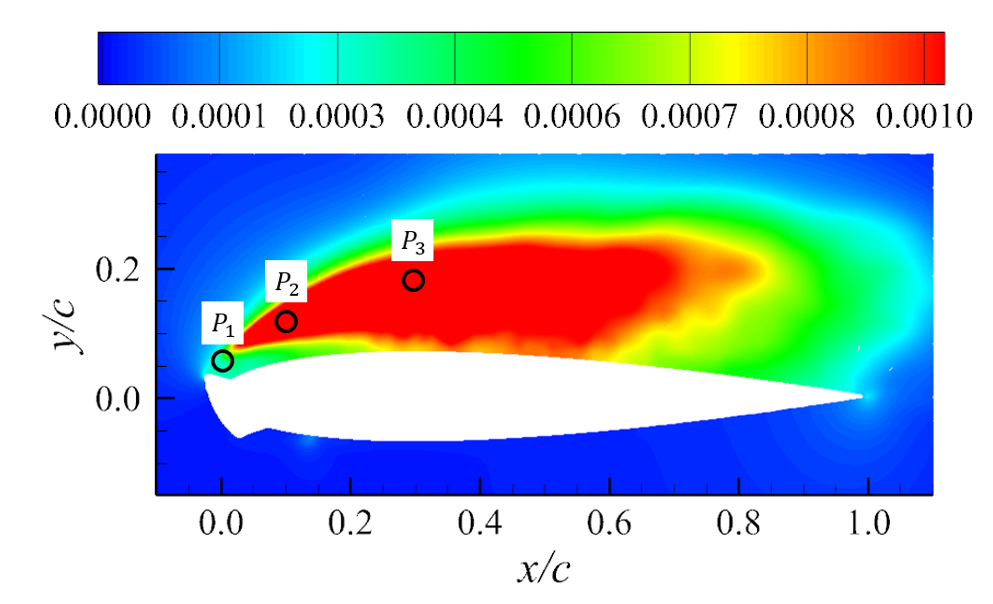}\hfill
  \caption{The sample points for spectral analysis at $z/b=0.56$ for the iced swept wing. The contour is the root mean square of the pressure coefficient $C_{p,rms}$ obtained via AMD-IDDES. }
  \label{fig:P_location}
\end{figure}

All cross-sections in this section correspond to those previously analyzed for pressure coefficient distributions. Specifically, for the iced swept wing, the sections are perpendicular to the wing leading edge at $z/b = 0.27$, 0.42, 0.56, 0.72, and 0.89, respectively. For the iced straight wing, the sections are also perpendicular to the wing leading edge and located at $z/b = 0.17$, 0.34, 0.50, 0.66, and 0.85, respectively. For convenience, these slices are hereafter referred to as slice 1 through slice 5. At five different cross-sections, three points were selected along the shear layer, with the corresponding $x$-coordinates remaining consistent across each section at 0, 0.1, and 0.3. The $y$-coordinates were adjusted according to the specific positions of the shear layer. Figure \ref{fig:P_location} displays the three sample points on the third cross-section for the iced swept wing. The point locations of the iced straight and swept wings are similar, with both distributed along the separated shear layers at five different spanwise sections.\par
%
\begin{figure}[!t]
\centering
\subfloat[Slice 1.]{\includegraphics[width=0.32\columnwidth]{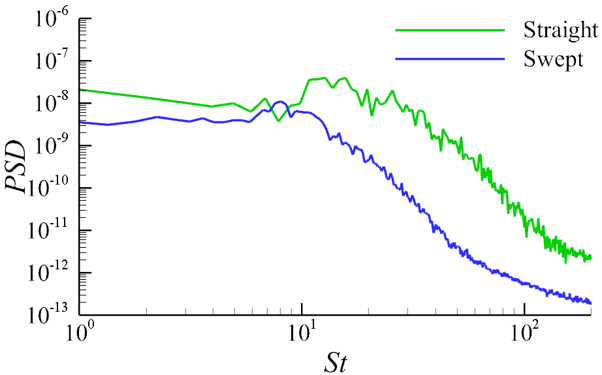}}%
\label{31c}
\hfil
\subfloat[Slice 2.]{\includegraphics[width=0.32\columnwidth]{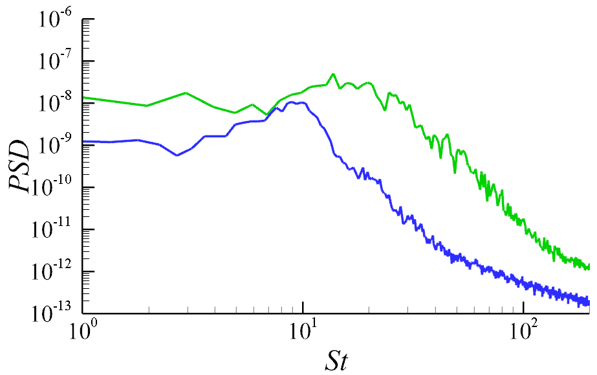}}%
\label{32c}
\subfloat[Slice 3.]{\includegraphics[width=0.32\columnwidth]{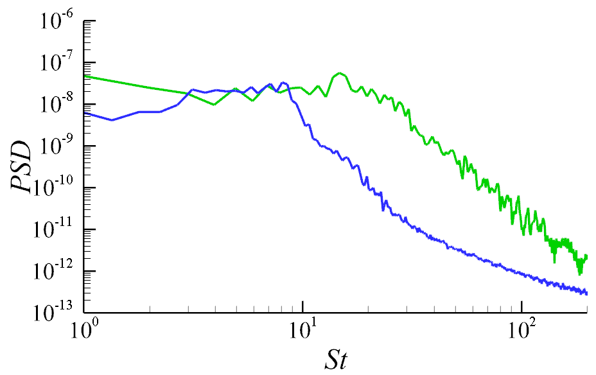}}%
\label{33c} \\
\subfloat[Slice 4.]{\includegraphics[width=0.32\columnwidth]{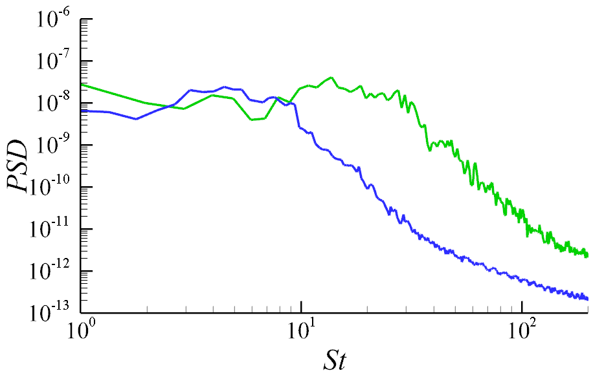}}%
\label{34c}
\subfloat[Slice 5.]{\includegraphics[width=0.32\columnwidth]{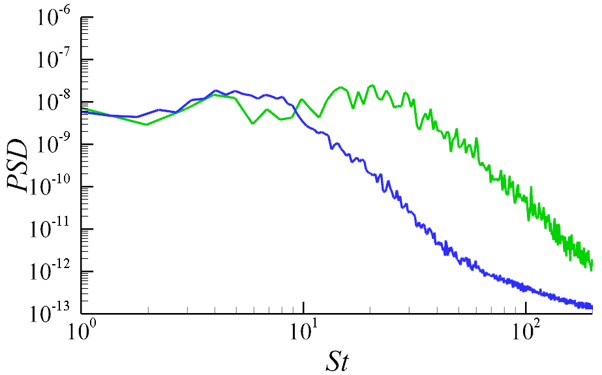}}
\label{35c}
\caption{Spectra of pressure fluctuations at $P_2$ at different spanwise locations for the iced straight wing.}
\label{fig:PSD3c}
\end{figure}

Figure \ref{fig:PSD3c} shows the spectra of pressure fluctuations at $P_2$ at different spanwise locations for the iced swept and straight wings. The vertical axis represents $\textit{PSD}$ $=(\Phi_{pp}/P_{\infty}^2)(U_{\infty}/c)$, while the horizontal axis represents the dimensionless frequency, the Strouhal number $St=fc/U_{\infty}$. At the $P_2$ position, a noticeable peak is observed for both the iced swept and straight wings. For the iced swept wing, the nondimensional peak frequency obtained by the AMD-IDDES method is approximately 9.5; for the iced straight wing, the corresponding value is 15, which is generally consistent with Xiao's results \cite{xiaoEnhancedPredictionThreedimensional2022}. Table \ref{tbl:st} presents the Strouhal numbers calculated at five spanwise sections (slice 1 to slice 5) for both the iced swept and straight wings, based on the vorticity thickness and local mean streamwise velocity:
\begin{equation}
    St _{\delta_{\varphi}}=\frac{f\delta_{\omega}}{\left\langle u\right\rangle_{\mathrm{ave}}}=\frac{f\delta_{\omega}}{\frac{\left\langle u\right\rangle_{\mathrm{max}}+\left\langle u\right\rangle_{\mathrm{min}}}{2}}
\end{equation}
\begin{equation}
    \delta_{\omega}=\frac{ \left\langle u \right\rangle_{max}-\left\langle u \right\rangle_{min}}{max\left( \frac{\partial\left\langle u \right\rangle}{\partial x} \right)}
\end{equation}
These frequencies are caused by the K--H instability, since they are close to the theoretical value of 0.143 for a classical mixing layer \cite{huerreLocalGlobalInstabilities2003}. The slightly higher current frequencies are reasonable, since the separating–reattaching flow differs from a canonical mixing layer due to reattachment. Richez et al. \cite{richezZonalDetachedEddySimulation2015} obtained $0.13-0.20$ for the shear stress layer emanating from a wing leading edge. For the iced straight wing and iced swept wing, the distribution of the peak Strouhal numbers at different sections generally agrees with the flow patterns analyzed in previous sections, further confirming the development of the shear layer along the spanwise direction in Figure \ref{fig:flow_model}.\par
%
\begin{table}[hbt!]
  \centering
  \caption{The peak Strouhal numbers in the initial shear stress layer at $P_2$.}
  \label{tbl:st}
  \begin{tabular}{ c c c c c c }
    \hline
    \hline
    \mc{${St}_{\delta_\omega}$}    & \mc{Slice 1} & \mc{Slice 2} & \mc{Slice 3} & \mc{Slice 4} & \mc{Slice 5}   \\
    \hline
    \small Iced straight wing  & $0.15$  & $0.25$  & $0.21$ & $0.19$  & $0.06$  \\
    \small Iced swept wing & $0.05$  & $0.18$ & $0.21$ & $0.33$  & $0.26$     \\
    \hline
    \hline
  \end{tabular}
\end{table}\par
Furthermore, the peak Strouhal numbers at position $P_2$ indicate that the corresponding frequencies on each section of the iced straight wing are higher than those of the iced swept wing. From the perspective of energy distribution, the spanwise pressure gradient on the swept wing drives the airflow along the spanwise direction, transferring part of the kinetic energy from the streamwise to the spanwise direction. This weakens the shedding capacity of streamwise vortices, thereby suppressing the strength and frequency of periodic shedding \cite{ribeiroTriglobalResolventAnalysis2023}. Meanwhile, the presence of a crossflow component within the boundary layer of the swept wing enhances its stability compared to a two-dimensional boundary layer, thereby inhibiting transition and the shedding of separation vortices. Additionally, the separated vortices on the swept wing are stretched along the spanwise direction, causing the vortex cores to be more susceptible to tearing and dissipation by the spanwise flow, further reducing the frequency of periodic shedding of streamwise vortices.\par
\begin{figure}[!t]
\centering
\subfloat[Slice 1.]{\includegraphics[width=0.32\columnwidth]{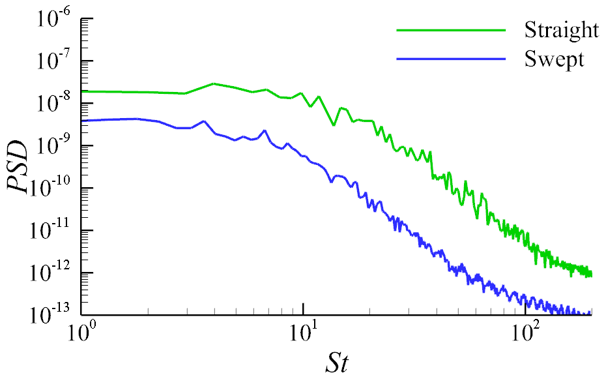}}%
\label{51c}
\hfil
\subfloat[Slice 3.]{\includegraphics[width=0.32\columnwidth]{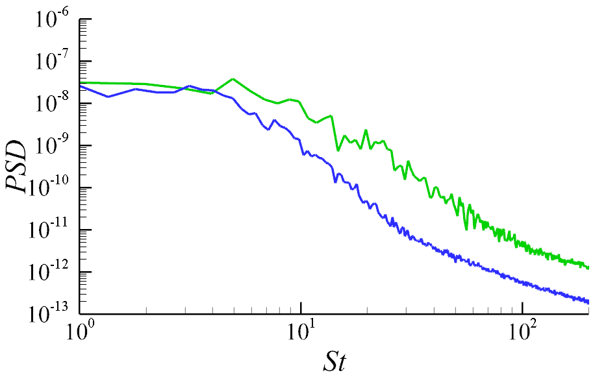}}%
\label{54c}
\subfloat[Slice 5.]{\includegraphics[width=0.32\columnwidth]{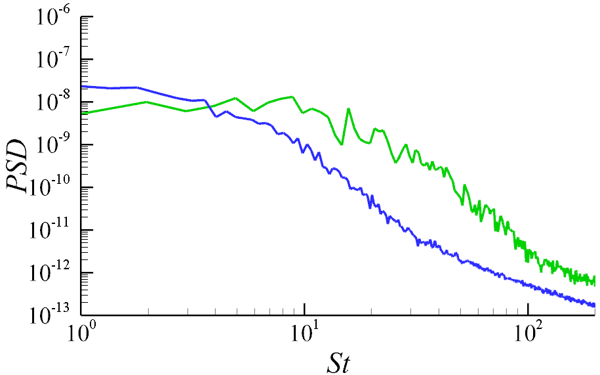}}
\label{55c}
\caption{The spectra of pressure fluctuations at $P_3$ at different spanwise locations for the iced straight wing.}
\label{fig:PSD5c}
\end{figure}

%
Figure \ref{fig:PSD5c} shows the pressure fluctuation spectrum at $P_3$. No clear peaks appear, indicating fully developed turbulence. The spectra across all sections are similar for the straight wing, with slightly lower low-frequency energy near the wingtip, suggesting uniform turbulence development and minimal spanwise variation.  In contrast, the swept wing shows spanwise variation: the low-frequency energy is much lower near the root due to delayed turbulence onset, then increases spanwise as the flow develops.  The spanwise flow promotes low-frequency energy growth, reflecting evolving turbulence structures.  The slightly reduced low-frequency energy near the wingtip is likely due to the wingtip vortex, which will be discussed later.\par

\begin{figure}[!t]
\centering
\subfloat[Slice 1.]{\includegraphics[width=0.32\columnwidth]{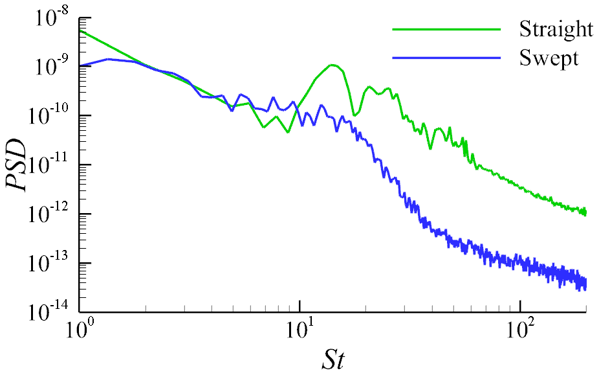}}%
\label{21c}
\hfil
\subfloat[Slice 2.]{\includegraphics[width=0.32\columnwidth]{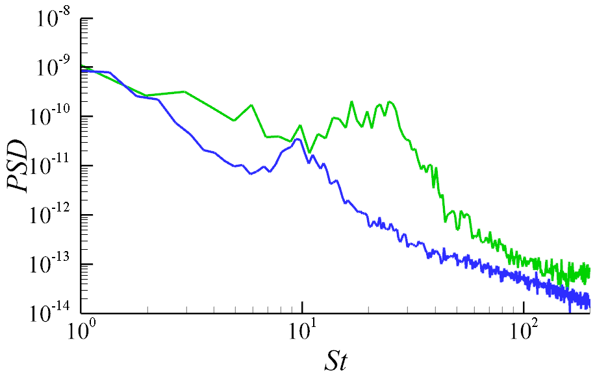}}%
\label{22c}
\subfloat[Slice 3.]{\includegraphics[width=0.32\columnwidth]{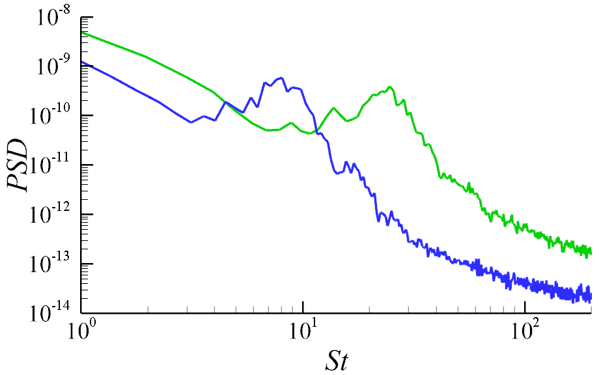}}%
\label{23c} \\
\subfloat[Slice 4.]{\includegraphics[width=0.32\columnwidth]{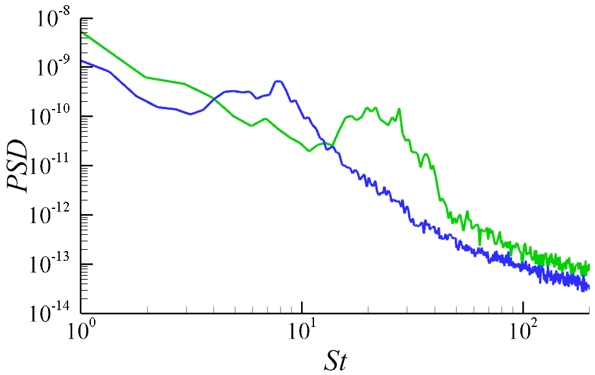}}%
\label{24c}
\subfloat[Slice 5.]{\includegraphics[width=0.32\columnwidth]{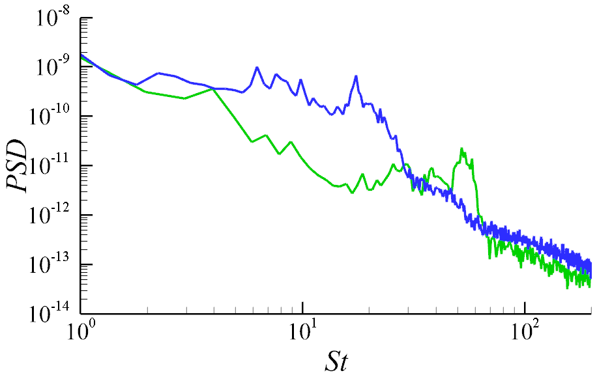}}
\label{25c}
\caption{Spectra of pressure fluctuations at $P_1$ at different spanwise locations for the iced straight wing.}
\label{fig:PSD2c}
\end{figure}

Figure \ref{fig:PSD2c} shows the spectra of the pressure fluctuations at $P_1$. The peak Strouhal number distributions across the five cross-sections of the iced straight wing and the iced swept wing exhibit similar trends: higher peak frequencies are observed in the mid-span region, with an increase near the wing root and wingtip. Notably, the peak Strouhal number near the wingtip is significantly more pronounced for the straight wing. These peak Strouhal numbers are relatively close to the peak Strouhal numbers at $P_2$ and still reflect the characteristics of shear layer separation. However, since $P_1$ is located just above the leading edge of the airfoil, very close to where the flow separates from the ice shape, the separation at $P_1$ is influenced not only by the shear layer behind the ice but also by additional effects near the wing root and wingtip. \par
In the wing root region, the peak Strouhal numbers of both the iced straight wing and the iced swept wing are higher than those in the mid-span region, with only a slight difference. This indicates that while the wing root region has an effect, this effect is not dominant. In the wingtip region, the peak Strouhal number is relatively high. This observation aligns with the conclusion drawn by Xiao et al. \cite{xiaoEnhancedPredictionThreedimensional2022} through their analysis of the spatial-temporal distributions of the $y$-velocity: the average time period of the vertical motion at the wingtip is shorter than that at the midspan. Theoretically, this phenomenon can be attributed to the interaction between the wingtip vortex and the leading-edge vortex. As previously discussed, the leading-edge vortex impedes the formation of the wingtip vortex. The wingtip vortex also suppresses the development of the leading-edge vortex. This occurs because the wingtip vortex reduces the freestream pressure in the vicinity of the wingtip, thereby reducing the pressure gradient necessary for the formation of the leading-edge vortex \cite{ribeiroTriglobalResolventAnalysis2023}. As a result, the wingtip vortex hinders the formation of the leading-edge vortex, weakening its strength and stability.\par
Next, a deeper analysis of the frequency differences between the two types of wings is conducted. The frequency differences corresponding to the peak values in the spectra of different sections of the swept wing are smaller than those of the straight wing. This is because the dominant flow in the iced swept wing is governed by the leading-edge vortex, with the wing root and wingtip having a relatively weaker influence on the statistical results. In contrast, the straight wing exhibits higher peak frequencies in the wingtip region, which further highlights the more pronounced effect of the wingtip vortex in the straight-wing flow, consistent with the results shown in Figure \ref{fig:vortx}.\par

\subsection{ Flow structure characterization}
POD is also called Karhunen--Loeve decomposition in the field of mathematics. It provides an algorithm to decompose a set of databases into a few basic modes with corresponding temporal coefficients and eigenvalues. POD enables the extraction of instantaneous flow data as high-resolution snapshots for further analysis. In this study, 400 sampling points were selected for eigenvalue analysis with the iced swept wing. The convergence of this dataset was previously validated by Refs. \cite{liuNumericalSimulationAerodynamic2023,zhangHighfidelityModelingTurbulent2022a}. The time-resolved flow fields were sampled with a nondimensional time interval of $\Delta t U_{\infty}/c = 0.05$. Due to the high computational cost, only five shedding cycles were included in the dataset. Nevertheless, this time window was sufficient to reveal the dominant flow structures, and a qualitative analysis of the periodic behavior in the modal amplitudes was conducted. \par 
The physical quantities commonly used in POD analysis include velocities and pressures. Preliminary investigations revealed that velocity-based POD fails to effectively capture the characteristics of shear layer separation and vortex evolution. This limitation is attributed to the strong three-dimensionality and spanwise development of the flow over the iced swept wing, which cannot be fully represented in the two-dimensional $x\text{-}y$ plane \cite{xiaoNumericalInvestigationUnsteady2020}. Therefore, the pressure is the target of the subsequent POD analysis, as it more clearly reveals the evolution of vortex structures, as discussed in the following sections. \par
\begin{figure}[!t]
\centering
\subfloat[Cumulative model contribution.]{\includegraphics[width=0.4\columnwidth]{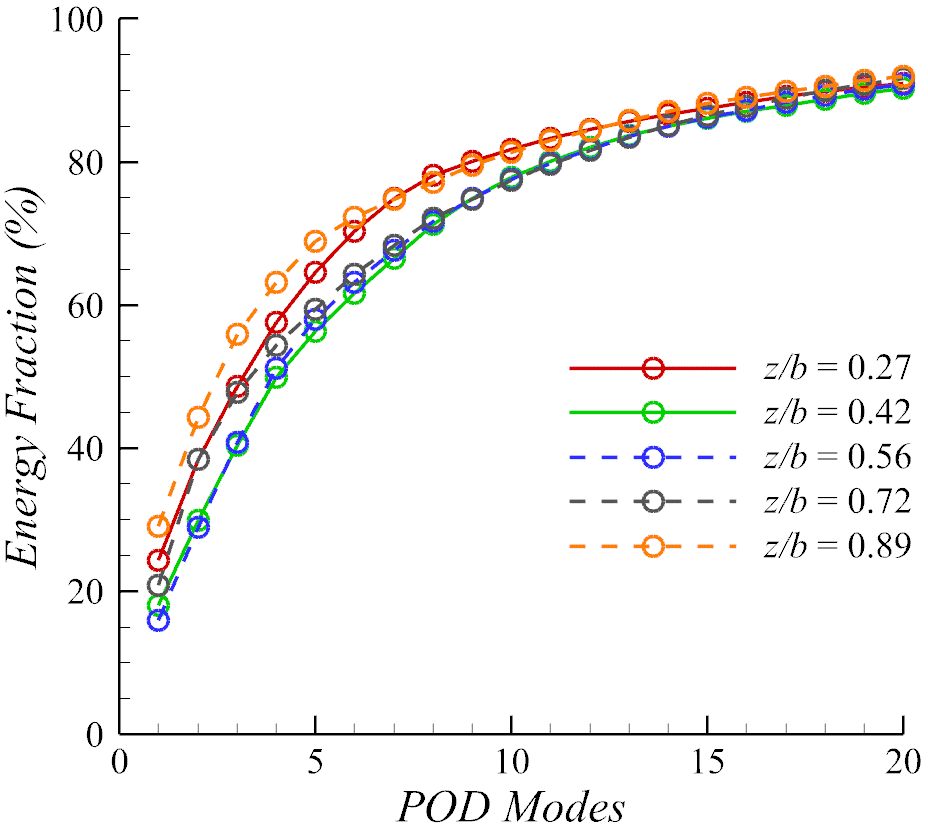}}%
\label{POD_CC} \hfil 
\subfloat[Energy fraction.]{\includegraphics[width=0.4\columnwidth]{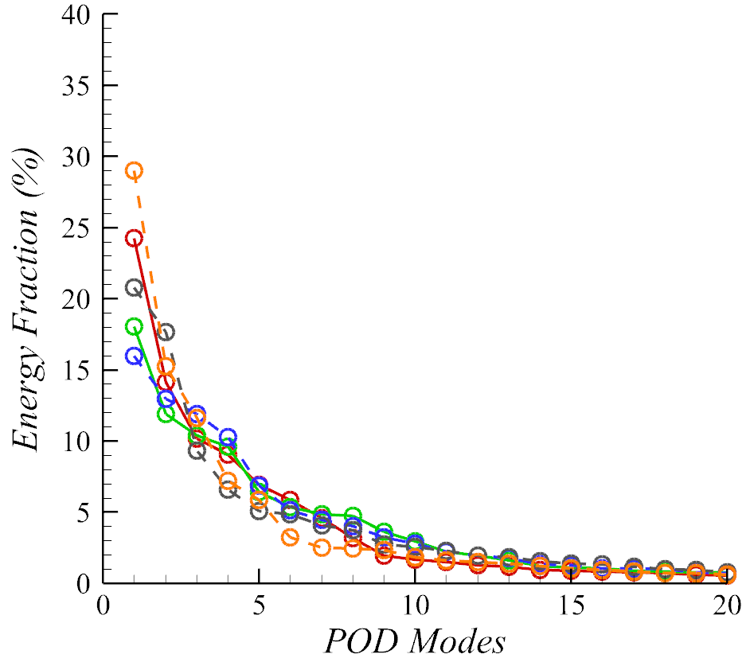}}%
\label{POD_EF} 
\caption{Cumulative model contribution and energy fraction of POD modes.}
\label{fig:POD_energy}
\end{figure}

Figure \ref{fig:POD_energy} and Table \ref{tbl:CC} present the cumulative model contribution and energy fraction of POD modes, which can serve as a criterion for selecting the dominant modes that capture most of the energy in the fluctuating flow field. The first 20 modes account for $90\%$ of the total energy in the flow field, while the first 50 modes capture up to $99\%$. The first four modes contribute the most, so the following analysis will primarily focus on these four dominant modes. \par
%
\begin{table}[hbt!]
  \centering
  \caption{Cumulative model contribution of POD modes.}
  \label{tbl:CC}
  \begin{tabular}{ c c c c c c }
    \hline
    \hline
    Slice    & Mode 1 & First five modes & First 10 modes & First 20 Modes & First 50 modes   \\
    \hline
    \small $z/b=0.27$ & $24\%$  & $64\%$ & $81\%$ & $91\%$  & $98\%$  \\
    \small $z/b=0.42$ & $18\%$  & $56\%$ & $77\%$ & $90\%$  & $98\%$  \\
    \small $z/b=0.56$ & $16\%$  & $58\%$ & $77\%$ & $91\%$  & $99\%$ \\
    \small $z/b=0.72$ & $21\%$  & $59\%$ & $77\%$ & $92\%$  & $99\%$ \\
    \small $z/b=0.89$ & $29\%$  & $69\%$ & $81\%$ & $92\%$  & $99\%$ \\
    \hline
    \hline
  \end{tabular}
\end{table}\par
\begin{figure}[!t]
\centering
\subfloat[Mode 1.]{\includegraphics[width=1.0\columnwidth]{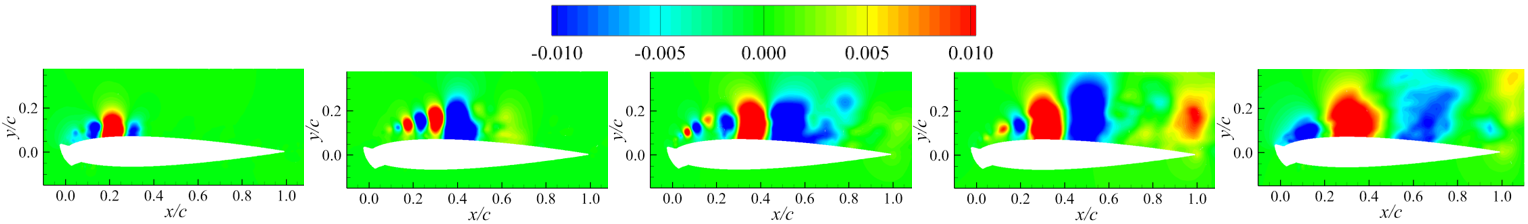}}%
\label{POD_1st} \hfil \\
\subfloat[Mode 2.]{\includegraphics[width=1.0\columnwidth]{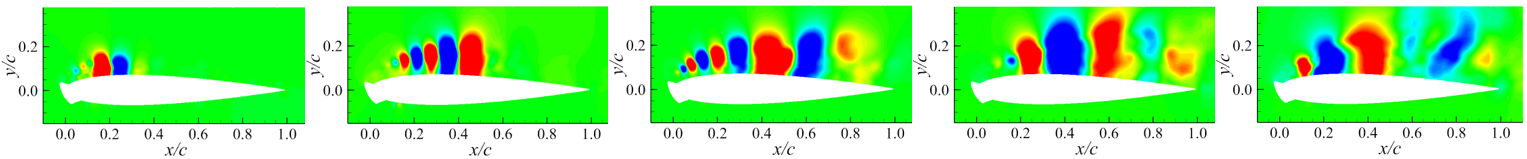}}%
\label{POD_2nd} \\
\subfloat[Mode 3.]{\includegraphics[width=1.0\columnwidth]{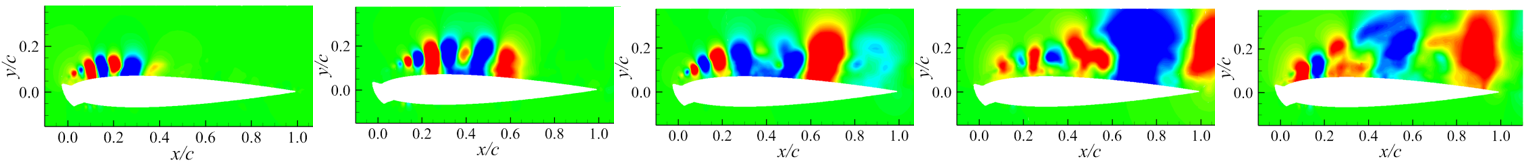}}%
\label{POD_3rd} \\
\subfloat[Mode 4.]{\includegraphics[width=1.0\columnwidth]{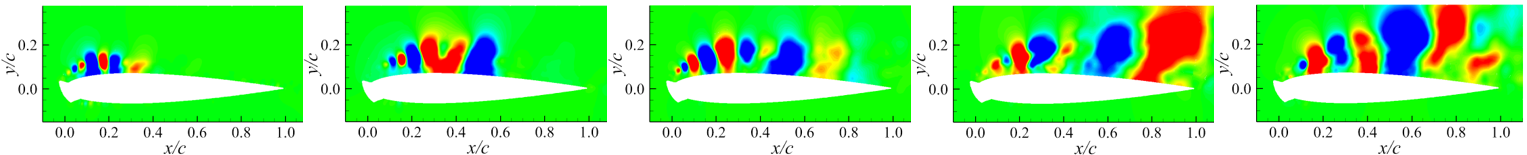}}%
\label{POD_4th} \\
\caption{The first four mode POD coefficient fields for different slices at $z/b = 0.27$, 0.42, 0.56, 0.72, and 0.89.}
\label{fig:POD_mode}
\end{figure}

Figure \ref{fig:POD_mode} shows the flow coefficient fields based on pressure fluctuations for the first four modes at different locations, as computed using the AMD-IDDES method. The dominant modes of the iced swept wing flow can be identified. The first and second modes exhibit a phase difference of about half a cycle, reflecting the dominant vortex patterns within the separation region. This clearly reveals the spanwise development trend of shear layer vortex shedding. The spatial structures of the third and fourth modes resemble those of the first two modes, where vortex pairing phenomena can be observed within the vortex structures \cite{xiaoNumericalInvestigationUnsteady2020}.\par
%
\begin{figure}[!t]
\centering
\subfloat[$z/b=0.27$.]{\includegraphics[width=0.32\columnwidth]{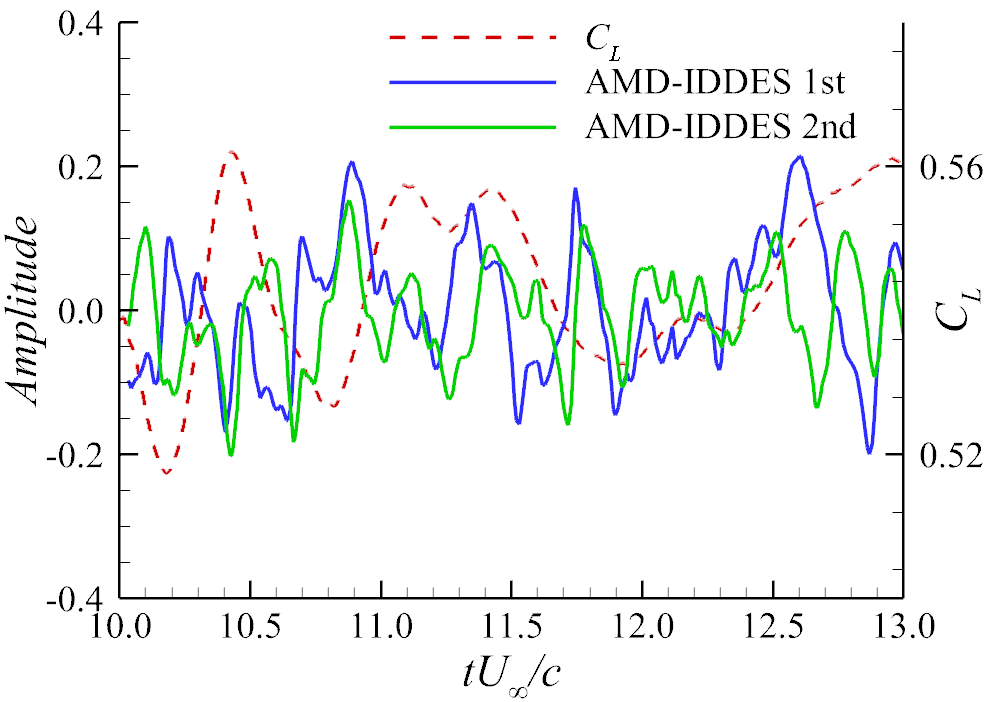}}%
\label{POD11}
\hfil
\subfloat[$z/b=0.42$.]{\includegraphics[width=0.32\columnwidth]{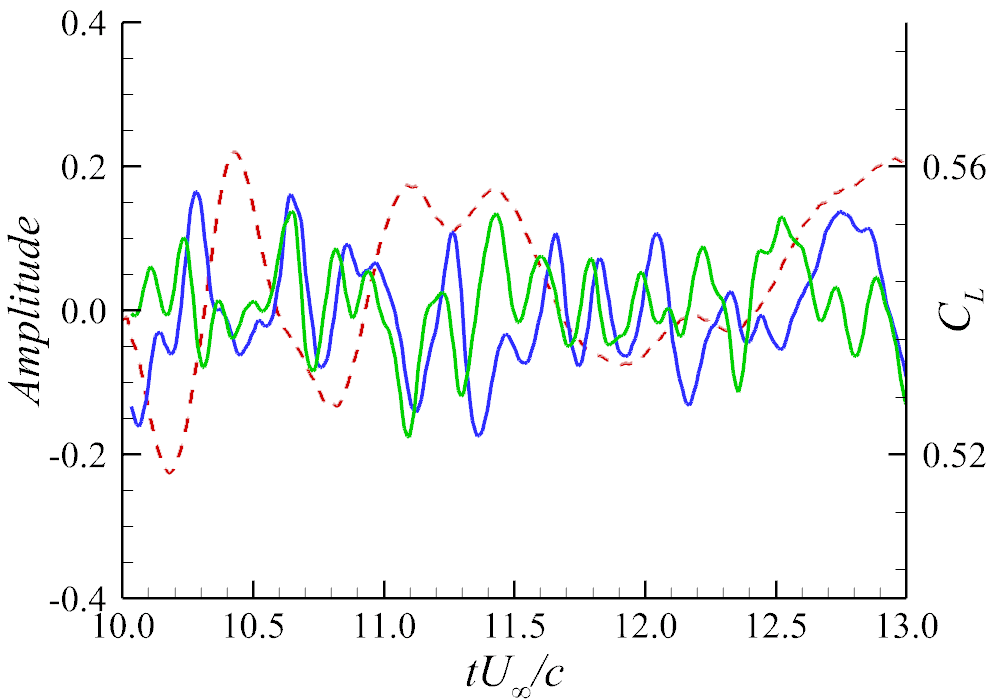}}%
\label{POD12}
\subfloat[$z/b=0.56$.]{\includegraphics[width=0.32\columnwidth]{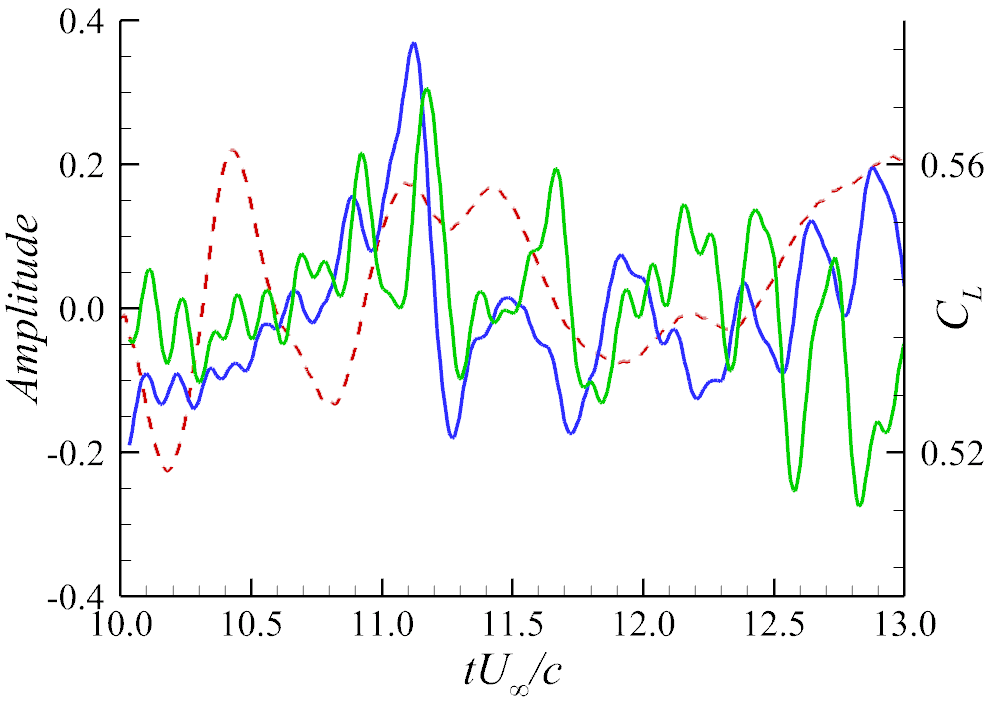}}%
\label{POD13} \\
\subfloat[$z/b=0.72$.]{\includegraphics[width=0.32\columnwidth]{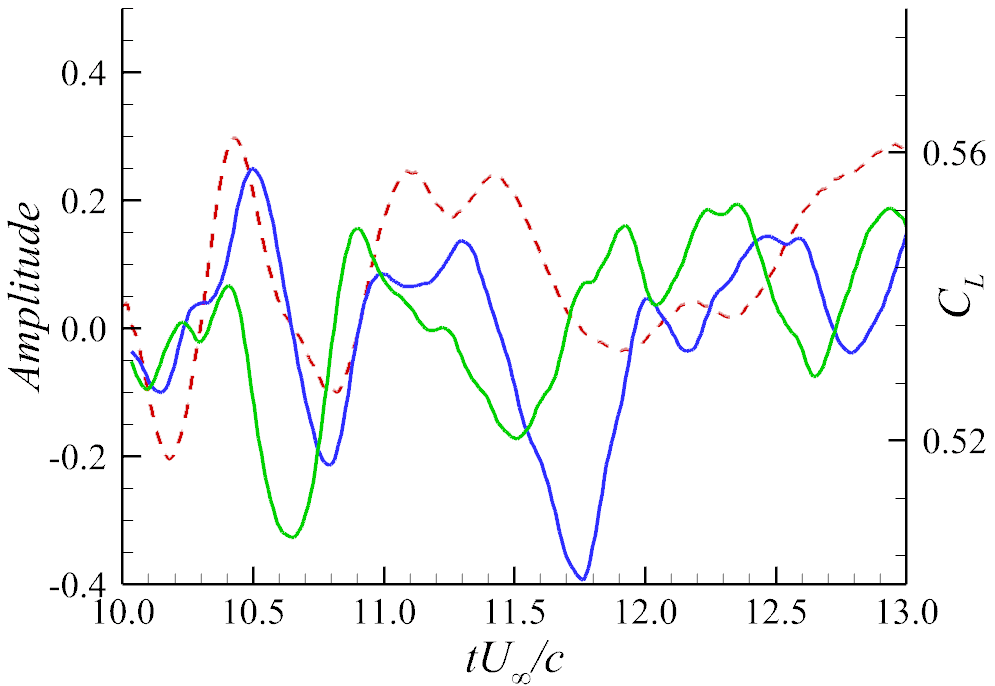}}%
\label{POD14}
\subfloat[$z/b=0.89$.]{\includegraphics[width=0.32\columnwidth]{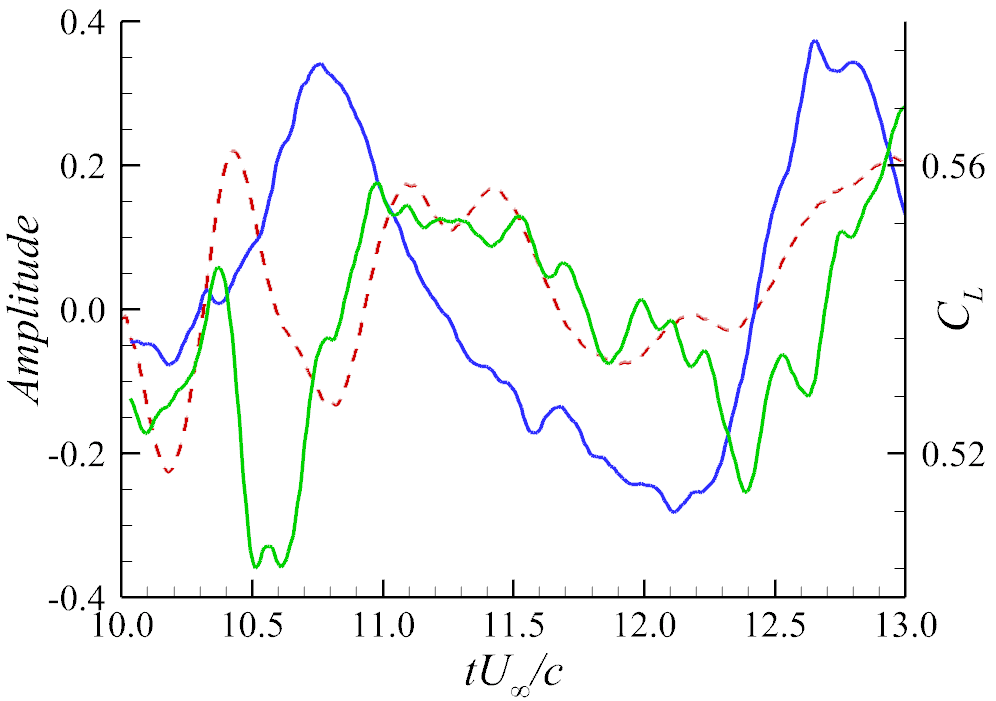}}%
\label{POD15} 
\caption{The time history of the lift force coefficient and the first and second temporal coefficients decomposed with POD for the iced swept wing.}
\label{fig:POD_1_2}
\end{figure}

\begin{figure}[!t]
\centering
\subfloat[$z/b=0.27$.]{\includegraphics[width=0.32\columnwidth]{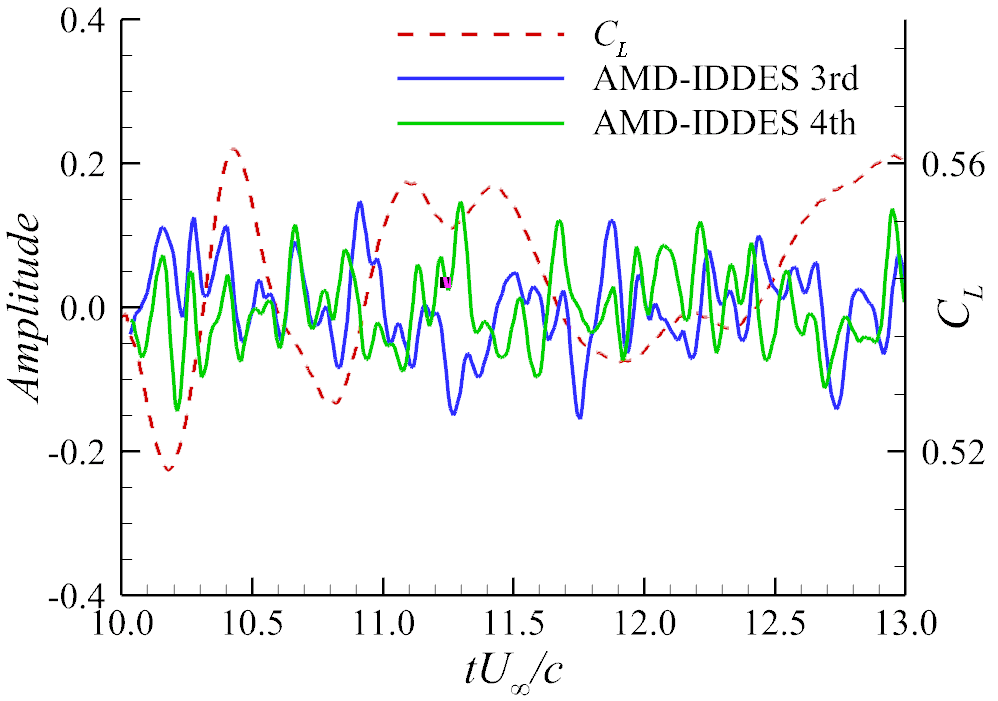}}%
\label{POD21}
\hfil
\subfloat[$z/b=0.42$.]{\includegraphics[width=0.32\columnwidth]{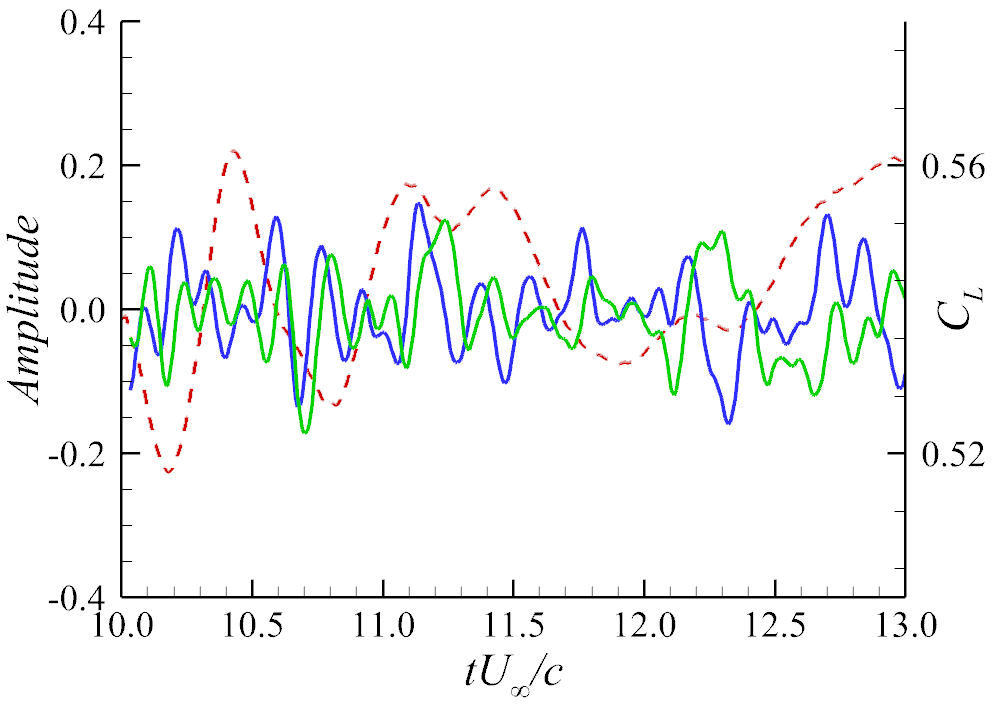}}%
\label{POD22}
\subfloat[$z/b=0.56$.]{\includegraphics[width=0.32\columnwidth]{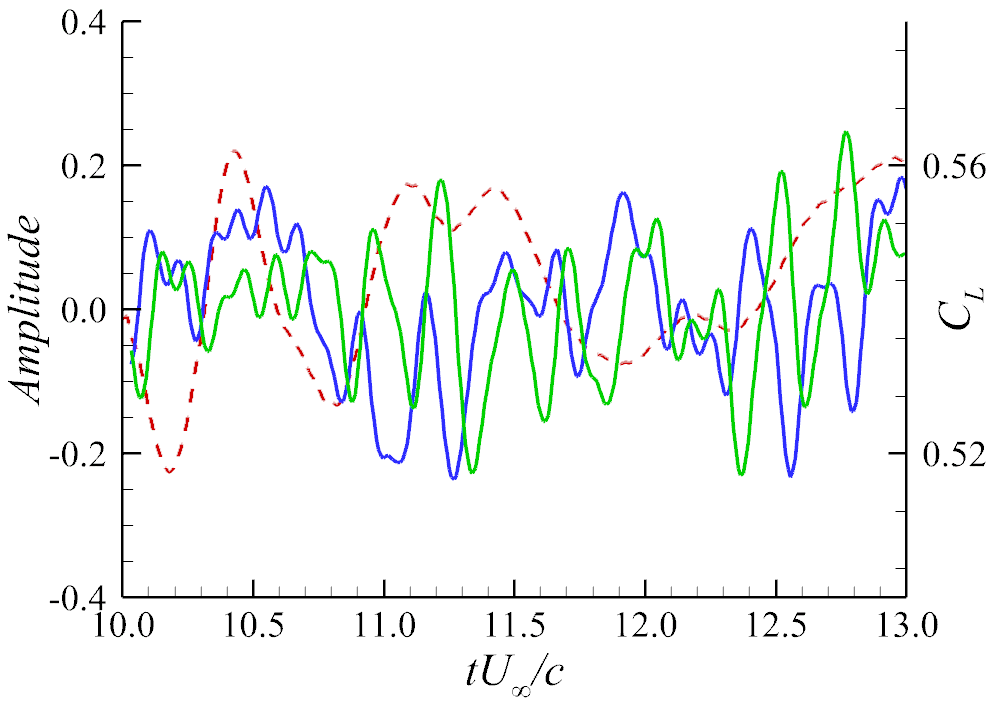}}%
\label{POD23} \\
\subfloat[$z/b=0.72$.]{\includegraphics[width=0.32\columnwidth]{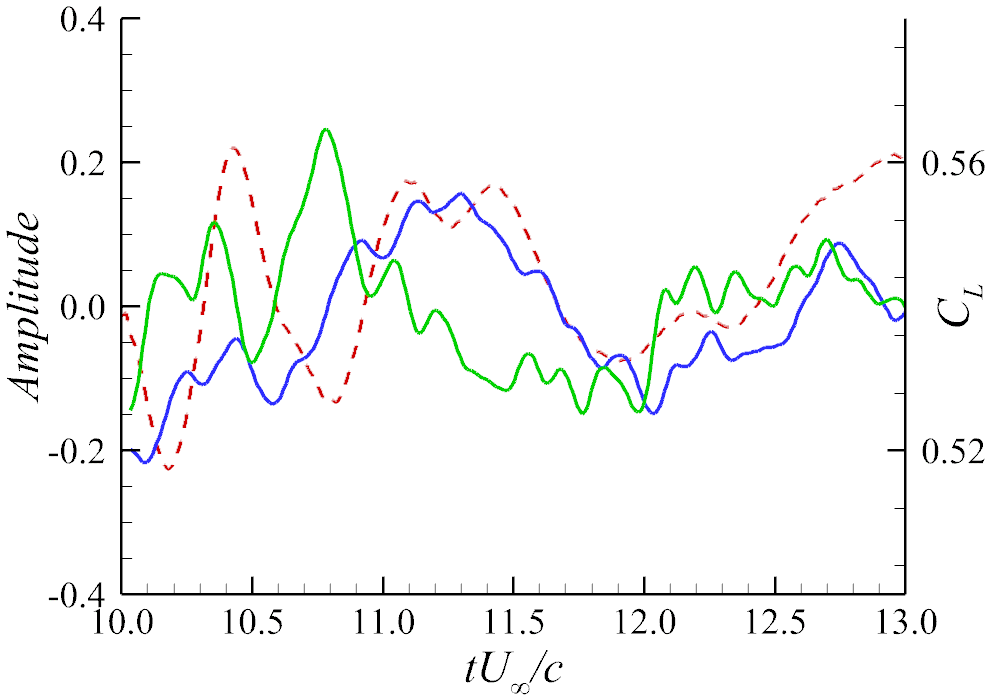}}%
\label{POD24}
\subfloat[$z/b=0.89$.]{\includegraphics[width=0.32\columnwidth]{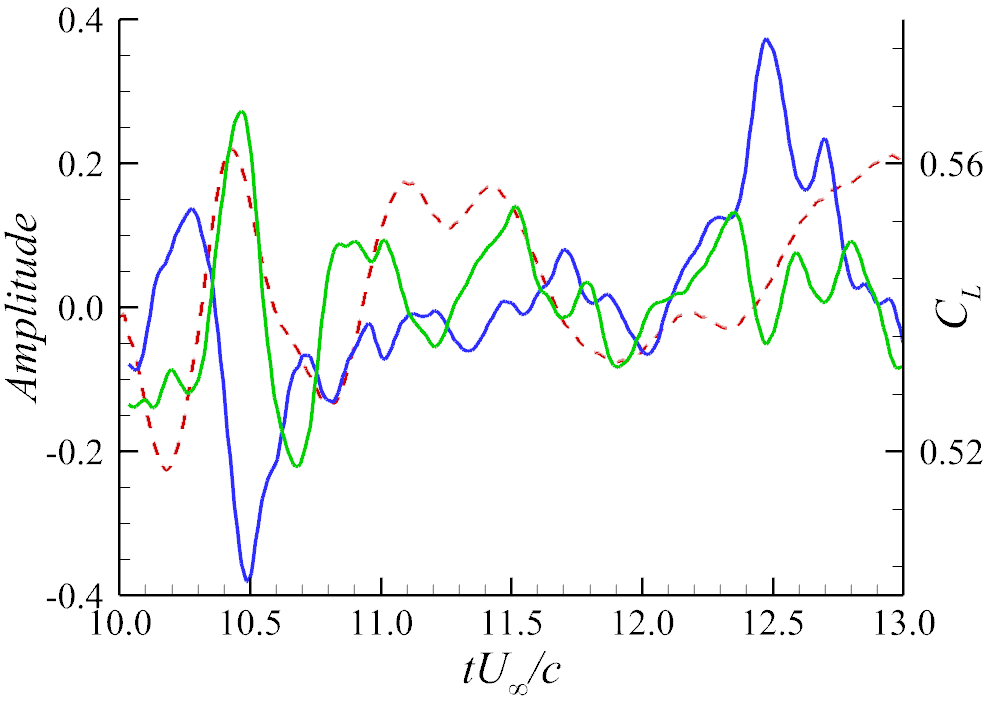}}%
\label{POD25} 
\caption{The time history of the lift force coefficient and the third and fourth temporal coefficients decomposed with POD for the iced swept wing.}
\label{fig:POD_3_4}
\end{figure}

Figures \ref{fig:POD_1_2} and \ref{fig:POD_3_4} show the time histories of the first four temporal coefficients obtained through POD decomposition and the lift coefficient of the iced swept wing. This comparison focuses solely on the time evolution of the temporal coefficients and the aerodynamic force coefficient, rather than analyzing the specific amplitudes \cite{liuNumericalSimulationAerodynamic2023}. A key observation is that among the five analyzed sections, the first-mode temporal coefficient at $z/b=0.72$ shows the strongest resemblance to the lift coefficient in both frequency and phase.  This indicates that the unsteady lift is primarily governed by the flow dynamics in the outboard region.  In this area, the leading-edge shear layer vortex has had sufficient space to evolve and stabilize, with minimal influence from the wing root or tip. As a result, large-scale, low-frequency separation emerges, which dominates the lift fluctuations.  This conclusion is well supported by the flow characteristics: the outboard region exhibits extensive flow separation that generates low-frequency oscillations.  Additionally, the POD spatial modes shown in Figure \ref{fig:POD_mode} clearly display dominant low-frequency, large-scale structures in the outboard part of the wing. \par
In contrast, the temporal coefficients near the wingtip display shorter oscillation periods and a weaker correlation with the lift fluctuations, suggesting a diminished role in driving unsteady aerodynamic forces.  This trend is consistent with the POD spatial modes, which reveal smaller-scale vortical structures in the tip region. Furthermore, the third and fourth temporal coefficients, corresponding to higher-order modes, exhibit higher-frequency oscillations and are associated with finer-scale, less coherent flow structures. These modes likely reflect background turbulence or acoustic disturbances and contribute little to the primary unsteady aerodynamic behavior.\par
\section{Conclusion}\label{sec:conclusions}
The aim of this paper was to investigate the flow characteristics of an iced swept wing using an improved delayed detached eddy simulation method, the AMD-IDDES method. To validate the reliability of the AMD-IDDES method, a comparison with the standard IDDES approach was first conducted to demonstrate its improvements, particularly in refining the subgrid length scale. These improvements reduce the eddy viscosity in the initial shear layer separation region, which accelerates the development of the Kelvin--Helmholtz (K--H) instability, leading to more accurate turbulence modeling. While the difference in pressure coefficient distributions between AMD-IDDES and IDDES is not significant when compared to experimental data, the AMD-IDDES method provides superior resolution of vortex structures and separated shear layers, proving its capability for accuracy in turbulent flow simulations.\par
The primary focus of this study was the analysis of the iced swept wing, which is characterized by the formation of a leading-edge vortex exhibiting strong spanwise flow. This spanwise motion significantly alters the separation and reattachment processes, distinguishing the aerodynamic behavior of the swept wing from that of a straight wing. The AMD-IDDES method effectively captures the unsteady evolution of these vortex structures, which originate from two-dimensional Kelvin–Helmholtz instabilities and transition into three-dimensional vortices. A wingtip vortex is also observed, although its influence is less dominant.\par
Unlike the streamwise flow development observed in the iced straight wing, the spanwise-evolving vortex structures in the iced swept wing profoundly influence the overall flow field. The spanwise flow induces a stall-like behavior in the wingtip region, reducing the pressure difference between the upper and lower surfaces. The interaction between the wingtip vortex and the leading-edge vortex not only affects the local flow near the wingtip and its downstream evolution but also modulates the frequency of shear layer separation. Further analysis reveals that the aerodynamic force coefficients are primarily governed by the characteristics of shear layer separation, particularly the dynamics of the leading-edge vortex. These features are closely associated with the flow near the mid-span region, while the wing root and wingtip effects on the overall aerodynamic response are negligible.\par
In conclusion, the AMD-IDDES method can successfully capture the complex flow structures and vortex dynamics over an iced swept wing. Through detailed analysis, the influence of the sweep angle on the flow characteristics of iced swept-wing configurations has been revealed, providing valuable insights into the aerodynamic impact of icing. These findings suggest that the AMD-IDDES method holds strong potential for broader application to more complex geometries and flow conditions in future studies of icing-related aerodynamic phenomena.\par

\section{Acknowledgments}\label{sec:Acknowledgments}
This work was supported by the National Natural Science Foundation of China (Grant Nos. 12388101, 12372288, U23A2069, and 92152301), the Jilin Province Science and Technology Development Program, China (Grant No.20220301013GX) and the National High Technology Research and Development Program of China (2024YFB4205601 and 2025YFB3003401). The authors would like to express their gratitude to Shaoguang Zhang and Haoyan Li for their insightful discussion. Special thanks are extended to Wybe Rozema for his helpful suggestions and discussions.

\section{Conflicts of interest}\label{sec:Conflicts of interest}
The authors report no conflicts of interest.
\section{Data availability statement}\label{sec:Data availability statement}
The data that support the findings of this study are available within the article.

\bibliography{sample}

\begin{thebibliography}{39}
\newcommand{\enquote}[1]{``#1''}
\providecommand{\natexlab}[1]{#1}
\providecommand{\url}[1]{\texttt{#1}}
\providecommand{\urlprefix}{URL }
\expandafter\ifx\csname urlstyle\endcsname\relax
  \providecommand{\doi}[1]{\discretionary{}{}{}https://doi.org/#1}\else
  \providecommand{\doi}[1]{\discretionary{}{}{}\urlstyle{rm}\url{https://doi.org/#1}}\fi

\bibitem[{Li et~al.(2021)Li, Zhang, and Chen}]{liNumericalSimulationIced2021}
Li, H., Zhang, Y., and Chen, H., \enquote{Numerical {{Simulation}} of {{Iced
  Wing Using Separating Shear Layer Fixed Turbulence Models}},} \emph{AIAA
  Journal}, Vol.~59, No.~9, 2021, pp. 3667--3681.
\newblock \doi{10.2514/1.J060143}.

\bibitem[{Bragg et~al.(2005)Bragg, Broeren, and
  Stirling}]{braggIceairfoilAerodynamics2005}
Bragg, M., Broeren, A., and Stirling, L., \enquote{Ice-Airfoil Aerodynamics,}
  \emph{Progress in Aerospace Sciences - PROG AEROSP SCI}, Vol.~41, 2005, pp.
  323--362.
\newblock \doi{10.1016/j.paerosci.2005.07.001}.

\bibitem[{Stebbins et~al.(2019)Stebbins, Loth, Broeren, and
  Potapczuk}]{stebbinsReviewComputationalMethods2019a}
Stebbins, S., Loth, E., Broeren, A., and Potapczuk, M., \enquote{Review of
  Computational Methods for Aerodynamic Analysis of Iced Lifting Surfaces,}
  \emph{Progress in Aerospace Sciences}, Vol. 111, 2019, p. 100583.
\newblock \doi{10.1016/j.paerosci.2019.100583}.

\bibitem[{Xiao et~al.(2022)Xiao, Zhang, and
  Zhou}]{xiaoEnhancedPredictionThreedimensional2022}
Xiao, M., Zhang, Y., and Zhou, F., \enquote{Enhanced Prediction of
  Three-Dimensional Finite Iced Wing Separated Flow near Stall,}
  \emph{International Journal of Heat and Fluid Flow}, Vol.~98, 2022, p.
  109067.
\newblock \doi{10.1016/j.ijheatfluidflow.2022.109067}.

\bibitem[{Bragg et~al.(2007)Bragg, Broeren, Addy, Potapczuk, Guffond, and
  Montreuil}]{braggAirfoilIceAccretionAerodynamic2007}
Bragg, M., Broeren, A., Addy, H., Potapczuk, M., Guffond, D., and Montreuil,
  E., \enquote{Airfoil {{Ice-Accretion Aerodynamic Simulation}},} \emph{45th
  {{AIAA Aerospace Sciences Meeting}} and {{Exhibit}}}, {American Institute of
  Aeronautics and Astronautics}, Reno, Nevada, 2007.
\newblock \doi{10.2514/6.2007-85}.

\bibitem[{Chen et~al.(2024)Chen, Zhang, and Fu}]{chenNumericalStudyEffects2024}
Chen, J., Zhang, Y., and Fu, S., \enquote{Numerical Study of the Effects of
  Discontinuous Ice on Three-Dimensional Wings,} \emph{Physics of Fluids},
  Vol.~36, No.~7, 2024, p. 075109.
\newblock \doi{10.1063/5.0217642}.

\bibitem[{Ansell and Bragg(2014)}]{ansellMeasurementUnsteadyFlow2014}
Ansell, P.~J., and Bragg, M.~B., \enquote{Measurement of {{Unsteady Flow
  Reattachment}} on an {{Airfoil}} with an {{Ice Shape}},} \emph{AIAA Journal},
  Vol.~52, No.~3, 2014, pp. 656--659.
\newblock \doi{10.2514/1.J052519}.

\bibitem[{Ozcer et~al.(2023)Ozcer, Pueyo, Menter, Hafid, and
  Yang}]{ozcerNumericalStudyIced2023a}
Ozcer, I., Pueyo, A., Menter, F., Hafid, S., and Yang, H., \enquote{Numerical
  {{Study}} of {{Iced Swept-Wing Performance Degradation}} Using {{RANS}},}
  \emph{International {{Conference}} on {{Icing}} of {{Aircraft}}, {{Engines}},
  and {{Structures}}}, Vienna, Austria, 2023, pp. 2023--01--1402.
\newblock \doi{10.4271/2023-01-1402}.

\bibitem[{Zhang et~al.(2015)Zhang, Habashi, and
  Khurram}]{zhangZonalDetachedEddySimulation2015}
Zhang, Y., Habashi, W., and Khurram, R., \enquote{Zonal {{Detached-Eddy
  Simulation}} of {{Turbulent Unsteady Flow}} over {{Iced Airfoils}},}
  \emph{Journal of Aircraft}, Vol.~53, 2015, pp. 1--14.
\newblock \doi{10.2514/1.C033253}.

\bibitem[{Molina et~al.(2020)Molina, Silva, Broeren, Righi, and
  Alonso}]{molinaApplicationDDESIced2020a}
Molina, E.~S., Silva, D.~M., Broeren, A.~P., Righi, M., and Alonso, J.~J.,
  \enquote{Application of {{DDES}} to {{Iced Airfoil}} in {{Stanford University
  Unstructured}} ({{SU2}}),} \emph{Progress in {{Hybrid RANS-LES Modelling}}},
  Vol. 143, edited by Y.~Hoarau, S.-H. Peng, D.~Schwamborn, A.~Revell, and
  C.~Mockett, Springer International Publishing, Cham, 2020, pp. 283--293.
\newblock \doi{10.1007/978-3-030-27607-2_23}.

\bibitem[{Xiao and Zhang(2021)}]{xiaoImprovedPredictionFlow2021}
Xiao, M., and Zhang, Y., \enquote{Improved {{Prediction}} of {{Flow Around
  Airfoil Accreted}} with {{Horn}} or {{Ridge Ice}},} \emph{AIAA Journal},
  Vol.~59, No.~6, 2021, pp. 2318--2327.
\newblock \doi{10.2514/1.J059744}.

\bibitem[{Butler et~al.(2016)Butler, Qin, and
  Loth}]{butlerImprovedDelayedDetachedEddy2016}
Butler, C., Qin, C., and Loth, E., \enquote{Improved {{Delayed Detached-Eddy
  Simulation}} on a {{Swept Hybrid Model}} in {{IRT}},} \emph{8th {{AIAA
  Atmospheric}} and {{Space Environments Conference}}}, {American Institute of
  Aeronautics and Astronautics}, Washington, D.C., 2016.
\newblock \doi{10.2514/6.2016-3736}.

\bibitem[{Xiao et~al.(2019)Xiao, Zhang, and Zhou}]{xiaoNumericalStudyIced2019}
Xiao, M., Zhang, Y., and Zhou, F., \enquote{Numerical {{Study}} of {{Iced
  Airfoils}} with {{Horn Features Using Large-Eddy Simulation}},} \emph{Journal
  of Aircraft}, Vol.~56, No.~1, 2019, pp. 94--107.
\newblock \doi{10.2514/1.C034986}.

\bibitem[{Xiao et~al.(2020)Xiao, Zhang, and
  Zhou}]{xiaoNumericalInvestigationUnsteady2020}
Xiao, M., Zhang, Y., and Zhou, F., \enquote{Numerical {{Investigation}} of the
  {{Unsteady Flow Past}} an {{Iced Multi-Element Airfoil}},} \emph{AIAA
  Journal}, Vol.~58, No.~9, 2020, pp. 3848--3862.
\newblock \doi{10.2514/1.J059114}.

\bibitem[{Lee et~al.(2020)Lee, Lee, Raj~L, Jo, and
  Myong}]{leeLargeeddySimulationsComplex2020}
Lee, Y., Lee, J.~H., Raj~L, P., Jo, J., and Myong, R.~S., \enquote{Large-Eddy
  Simulations of Complex Aerodynamic Flows over Multi-Element Iced Airfoils,}
  \emph{Aerospace Science and Technology}, Vol. 109, 2020, p. 106417.
\newblock \doi{10.1016/j.ast.2020.106417}.

\bibitem[{Broeren et~al.(2017)Broeren, Woodard, Diebold, and
  Moens}]{broerenLowReynoldsNumberAerodynamics2017a}
Broeren, A.~P., Woodard, B., Diebold, J.~M., and Moens, F.,
  \enquote{Low-Reynolds Number Aerodynamics of an 8.9\% Scale Semispan Swept
  Wing for Assessment of Icing Effects,} \emph{9th AIAA Atmospheric and Space
  Environments Conference}, American Institute of Aeronautics and Astronautics,
  Denver, Colorado, 2017.
\newblock \doi{10.2514/6.2017-4372}.

\bibitem[{Bornhoft et~al.(2024)Bornhoft, Jain, Bose, and
  Moin}]{bornhoftLargeeddySimulationsCRM652024a}
Bornhoft, B.~J., Jain, S.~S., Bose, S.~T., and Moin, P., \enquote{Large-Eddy
  Simulations of the {{CRM65}} Swept Wing under Real and Artificial Icing
  Conditions,} \emph{{{AIAA SCITECH}} 2024 {{Forum}}}, {American Institute of
  Aeronautics and Astronautics}, Orlando, FL, 2024.
\newblock \doi{10.2514/6.2024-1336}.

\bibitem[{Bornhoft et~al.(2025)Bornhoft, Moin, Jain, and
  Bose}]{bornhoftUseArtificialIce2025}
Bornhoft, B., Moin, P., Jain, S.~S., and Bose, S.~T., \enquote{On the {{Use}}
  of {{Artificial Ice Shapes}} for {{Large-Eddy Simulations}} in {{Aircraft
  Icing}},} \emph{Journal of Aircraft}, 2025, pp. 1--18.
\newblock \doi{10.2514/1.C038146}.

\bibitem[{Stebbins and Loth(2024)}]{stebbinsNumericalSimulationIced2024}
Stebbins, S., and Loth, E., \enquote{Numerical Simulation of Iced Swept Wing
  Aerodynamics with {{RANS}}, {{DES}}, and {{IDDES}},} \emph{Handbook of
  {{Numerical Simulation}} of {{In-Flight Icing}}}, edited by W.~G. Habashi,
  Springer International Publishing, Cham, 2024, pp. 481--502.
\newblock \doi{10.1007/978-3-031-33845-8_6}.

\bibitem[{Craig~Penner et~al.(2024)Craig~Penner, Housman, Stich, Sousa, Koch,
  and Duensing}]{craigpennerWallModeledLargeEddySimulations2024b}
Craig~Penner, D., Housman, J.~A., Stich, G.-D., Sousa, V.~C., Koch, J.~R., and
  Duensing, J., \enquote{Wall-{{Modeled Large-Eddy Simulations}} of a {{Swept
  Wing With Leading-Edge Ice}},} \emph{{{AIAA AVIATION FORUM AND ASCEND}}
  2024}, {American Institute of Aeronautics and Astronautics}, Las Vegas,
  Nevada, 2024.
\newblock \doi{10.2514/6.2024-3999}.

\bibitem[{Craig~Penner et~al.(2025)Craig~Penner, Sousa, Ravikumar, Saurabh,
  Cadieux, Stich, Barad, Housman, and
  Duensing}]{craigpennerWallModeledSweptWing2025b}
Craig~Penner, D., Sousa, V., Ravikumar, K., Saurabh, K., Cadieux, F., Stich,
  G.-D., Barad, M.~F., Housman, J.~A., and Duensing, J.~C.,
  \enquote{Wall-{{Modeled LES}} of a {{Swept Wing With Leading-Edge Ice Using
  LAVA Curvilinear}}, {{Unstructured}}, and {{Cartesian Solvers}},}
  \emph{{{AIAA SCITECH}} 2025 {{Forum}}}, {American Institute of Aeronautics
  and Astronautics}, Orlando, FL, 2025.
\newblock \doi{10.2514/6.2025-0059}.

\bibitem[{Zhou et~al.(2025)Zhou, Xiao, Li, and
  Zhang}]{zhouEnhancedDelayedDetachededdy2025}
Zhou, Z., Xiao, M., Li, D., and Zhang, Y., \enquote{Enhanced Delayed
  Detached-Eddy Simulation with Anisotropic Minimum Dissipation Subgrid Length
  Scale,} \emph{Physics of Fluids}, Vol.~37, No.~2, 2025, p. 025152.
\newblock \doi{10.1063/5.0246596}.

\bibitem[{Rozema et~al.(2015)Rozema, Bae, Moin, and
  Verstappen}]{rozemaMinimumdissipationModelsLargeeddy2015}
Rozema, W., Bae, H.~J., Moin, P., and Verstappen, R.,
  \enquote{Minimum-Dissipation Models for Large-Eddy Simulation,} \emph{Physics
  of Fluids}, Vol.~27, 2015.
\newblock \doi{10.1063/1.4928700}.

\bibitem[{Gritskevich et~al.(2012)Gritskevich, Garbaruk, Sch{\"u}tze, and
  Menter}]{gritskevichDevelopmentDDESIDDES2012}
Gritskevich, M., Garbaruk, A., Sch{\"u}tze, J., and Menter, F.,
  \enquote{Development of {{DDES}} and {{IDDES Formulations}} for the
  K-{$\omega$} {{Shear Stress Transport Model}},} \emph{Flow, Turbulence and
  Combustion}, Vol.~88, 2012.
\newblock \doi{10.1007/s10494-011-9378-4}.

\bibitem[{{NASA Langley Research Center}(2017)}]{CFL3DVersion}
{NASA Langley Research Center}, \enquote{{CFL3D} Version 6.7,}
  \url{https://nasa.github.io/CFL3D/}, 2017.

\bibitem[{Kloker and Borsboom(1986)}]{klokerFullyImplicitScheme1986}
Kloker, M., and Borsboom, M., \enquote{A Fully Implicit Scheme for Unsteady
  Flow Calculations Solved by the Approximate Factorization Technique -
  Implicit Dual Time Stepping,} Tech. Rep. VKI Project Report 1986-16, VKI,
  Rhode St. Genese, Belgium, June 1986.

\bibitem[{Liu et~al.(2018)Liu, Zhu, Xiao, Sun, Huang, and
  Liu}]{liuDDESAdaptiveCoefficient2018}
Liu, J., Zhu, W., Xiao, Z., Sun, H., Huang, Y., and Liu, Z., \enquote{{{DDES}}
  with Adaptive Coefficient for Stalled Flows Past a Wind Turbine Airfoil,}
  \emph{Energy}, Vol. 161, 2018.
\newblock \doi{10.1016/j.energy.2018.07.176}.

\bibitem[{{Comte-Bellot} and
  Corrsin(1971)}]{comte-bellotSimpleEulerianTime1971}
{Comte-Bellot}, G., and Corrsin, S., \enquote{Simple {{Eulerian}} Time
  Correlation of Full-and Narrow-Band Velocity Signals in Grid-Generated,
  `Isotropic' Turbulence,} \emph{Journal of Fluid Mechanics}, Vol.~48, No.~2,
  1971, pp. 273--337.
\newblock \doi{10.1017/S0022112071001599}.

\bibitem[{Khodadoust and Bragg(1995)}]{khodadoustAerodynamicsFiniteWing1995}
Khodadoust, A., and Bragg, M.~B., \enquote{Aerodynamics of a Finite Wing with
  Simulated Ice,} \emph{Journal of Aircraft}, Vol.~32, No.~1, 1995, pp.
  137--144.
\newblock \doi{10.2514/3.46694}.

\bibitem[{Bragg et~al.(1991{\natexlab{a}})Bragg, Khodadoust, Soltani, Wells,
  and Kerho}]{braggEffectSimulatedIce1991}
Bragg, M., Khodadoust, A., Soltani, R., Wells, S., and Kerho, M.,
  \enquote{Effect of a Simulated Ice Accretion on the Aerodynamics of a Swept
  Wing,} \emph{29th {{Aerospace Sciences Meeting}}}, {American Institute of
  Aeronautics and Astronautics}, Reno,NV,U.S.A., 1991{\natexlab{a}}.
\newblock \doi{10.2514/6.1991-442}.

\bibitem[{Bragg et~al.(1991{\natexlab{b}})Bragg, Khodadoust, Soltani, Wells,
  and Kerho}]{braggAerodynamicMeasurementsFinite1991}
Bragg, M., Khodadoust, A., Soltani, R., Wells, S., and Kerho, M.,
  \enquote{Aerodynamic Measurements on a Finite Wing with Simulated Ice,}
  \emph{9th {{Applied Aerodynamics Conference}}}, {American Institute of
  Aeronautics and Astronautics}, Baltimore,MD,U.S.A., 1991{\natexlab{b}}.
\newblock \doi{10.2514/6.1991-3217}.

\bibitem[{Lee et~al.(2015)Lee, Kawai, Nonomura, Anyoji, Aono, Oyama, Asai, and
  Fujii}]{leeMechanismsSurfacePressure2015}
Lee, D., Kawai, S., Nonomura, T., Anyoji, M., Aono, H., Oyama, A., Asai, K.,
  and Fujii, K., \enquote{Mechanisms of Surface Pressure Distribution within a
  Laminar Separation Bubble at Different {{Reynolds}} Numbers,} \emph{Physics
  of Fluids}, Vol.~27, No.~2, 2015, p. 023602.
\newblock \doi{10.1063/1.4913500}.

\bibitem[{Deck and
  Thorigny(2007)}]{deckUnsteadinessAxisymmetricSeparatingreattaching2007}
Deck, S., and Thorigny, P., \enquote{Unsteadiness of an Axisymmetric
  Separating-Reattaching Flow: {{Numerical}} Investigation,} \emph{Physics of
  Fluids}, Vol.~19, No.~6, 2007, p. 065103.
\newblock \doi{10.1063/1.2734996}.

\bibitem[{Kwon and Sankar(1991)}]{kwonNumericalStudyEffects1991}
Kwon, O., and Sankar, L., \enquote{Numerical Study of the Effects of Icing on
  Fixed and Rotary Wing Performance,} \emph{AIAA Paper}, , No. 91-0662, 1991.

\bibitem[{Huerre and Monkewitz(2003)}]{huerreLocalGlobalInstabilities2003}
Huerre, P., and Monkewitz, P., \enquote{Local and {{Global Instabilities}} in
  {{Spatially Developing Flows}},} \emph{Annual Review of Fluid Mechanics},
  Vol.~22, 2003, pp. 473--537.
\newblock \doi{10.1146/annurev.fl.22.010190.002353}.

\bibitem[{Richez et~al.(2015)Richez, Pape, and
  Costes}]{richezZonalDetachedEddySimulation2015}
Richez, F., Pape, A., and Costes, M., \enquote{Zonal {{Detached-Eddy
  Simulation}} of {{Separated Flow Around}} a {{Finite-Span Wing}},} \emph{AIAA
  Journal}, Vol.~52, 2015, pp. 1--10.
\newblock \doi{10.2514/1.J053636}.

\bibitem[{Ribeiro et~al.(2023)Ribeiro, Yeh, and
  Taira}]{ribeiroTriglobalResolventAnalysis2023}
Ribeiro, J., Yeh, C.-A., and Taira, K., \enquote{Triglobal Resolvent Analysis
  of Swept-Wing Wakes,} \emph{Journal of Fluid Mechanics}, Vol. 954, 2023.
\newblock \doi{10.1017/jfm.2022.1033}.

\bibitem[{Liu and Zhang(2023)}]{liuNumericalSimulationAerodynamic2023}
Liu, H., and Zhang, C., \enquote{Numerical {{Simulation}} of {{Aerodynamic
  Features}} with {{Ice Shapes}} via {{High-Fidelity CFD Method}},}
  \emph{Handbook of {{Numerical Simulation}} of {{In-Flight Icing}}}, edited by
  W.~G. Habashi, Springer International Publishing, Cham, 2023, pp. 1--40.
\newblock \doi{10.1007/978-3-030-64725-4_45-1}.

\bibitem[{Zhang et~al.(2022)Zhang, Tan, Xu, Li, Fuxin, and
  Liu}]{zhangHighfidelityModelingTurbulent2022a}
Zhang, C., Tan, X., Xu, W., Li, G., Fuxin, W., and Liu, H.,
  \enquote{High-Fidelity Modeling of Turbulent Shear Flow Downstream of a
  3-{{D}} Airfoil with Spanwise Ice Contamination Leading Stall,}
  \emph{Computers \& Fluids}, Vol. 240, 2022, p. 105423.
\newblock \doi{10.1016/j.compfluid.2022.105423}.

\end{thebibliography}

\end{document}